\documentclass[onecolumn, usenatbib]{mnras}
\usepackage{savesym}
\usepackage{graphicx}
\usepackage{longtable}
\usepackage{changepage}

\usepackage{subfig}
\usepackage{amsmath}
\usepackage{amssymb}
\usepackage{verbatim}
\usepackage[yyyymmdd,hhmmss]{datetime}
\usepackage{array}
\usepackage{times}
\usepackage{xcolor}

\title[The X-ray variability of tidal disruption events]{ The X-ray variability of tidal disruption events }

\author [Andrew Mummery]{Andrew Mummery$^1$\thanks{E-mail:
andrew.mummery@physics.ox.ac.uk}\\
$^1$Oxford Theoretical Physics, Beecroft Building,  Clarendon Laboratory, Parks Road, Oxford, OX1 3PU, United Kingdom  }

\date{}
\label{firstpage}
\pagerange{\pageref{firstpage}--\pageref{lastpage}}

\begin{document}

\maketitle

\begin{abstract}
    When a star is torn apart by the tidal forces of a supermassive black hole (a so-called TDE) a transient accretion episode is initiated and a hot, often X-ray bright, accretion disk is formed. Like any accretion flow this disk is turbulent, and therefore the emission from its surface will vary stochastically. As the disk has a finite mass supply (i.e., at most the initial mass of the disrupted star) the disk will also undergo long-timescale evolution, as this material is lost into the black hole. In this paper we combine theoretical models for this long time evolution of the disk with models for the  stochastic variability of turbulent accretion flows which are correlated on short (orbital) timescales. This new framework allows us to demonstrate that (i) dimming events should be more prevalent than brightening events in long term TDE X-ray light curves (i.e., their log-luminosity distribution should be asymmetric), (ii) TDE X-ray light curves should follow a near- (but formally sub-)linear correlation between their root mean square variability and the mean flux, (iii) the fractional variability observed on short timescales across an X-ray observing band should increase with observing energy, and (iv) TDEs offer a unique probe of the physics of disk turbulence, owing to their clean spectra and natural evolutionary timescales. We confirm predictions (i) and (ii) with an analysis of the long timescale variability of two observed TDEs, and show strong support for prediction (iii) using the intra-observation variability of the same two sources. 
\end{abstract}

\section{Introduction}
Accretion disks are fundamentally turbulent systems, a result of the magneto-rotational instability \citep{BalbusHawley91}. As such, the observed emission from an accretion disk varies stochastically across a vast array of timescales.  This variability can be as rapid as the local orbital timescale of the very inner disk, or orders of magnitude longer than any physical process which is involved in the direct production of the disk's luminosity.  Broadly similar gross disk variability properties are also observed across different systems which span a huge array of length scales: from the compact disks in Galactic X-ray binaries \citep[e.g.,][]{Uttley01}, all the way up to the colossal disks in  active galactic nuclei \citep[e.g.,][]{Vaughan2011}. 

The soft X-ray light curves and spectra of many tidal disruption events (TDEs) are well-described as resulting from an evolving relativistic accretion disk \citep[e.g.,][]{MumBalb20a, Wen20, Guolo24, GuoloMum24}. The global evolutionary timescales of these disk systems are typically tens to hundreds of days, significantly longer than the timescales over which TDEs are observed in X-rays \citep[which can be as short as $\sim$ hours-to-days, e.g.,][among many others]{Wevers19b, Neustadt20, Pasham24, Yuhan24, Masterson25}
This means that these observations of TDEs probe rapid timescales well below the ``viscous'' time, and therefore significant deviations from smooth, diffusive, disk evolution may be expected.   

Understanding the properties of TDE X-ray light curve fluctuations will only grow in importance as the sample size of X-ray bright TDEs increases in the coming years. Indeed, the current sample of X-ray bright TDEs already appear to show significant X-ray variability, but due to sampling issues in follow-up observations (for example, non uniform observing cadences and big gaps between observations), such behavior is currently poorly characterized. Despite this, examples of different types of behavior already observed include rapid ($\sim$hours-weeks) flaring/dipping episodes \citep[e.g.,][]{Saxton12, Yuhan24}, sudden \citep{Kajava20} or smooth \citep{Gezari17} X-ray brightening with significant delays with respect to the UV/optical peak. Following such behavior, the X-ray emission may then rapidly fade away again over $\sim$ week timescales \citep{Liu23}. See e.g., Figures \ref{fig:14li_data} and \ref{fig:22lri_data} for two different examples of TDE X-ray variability analyzed in this work. 

It is interesting to note that soft TDE X-ray emission is significantly more variable than X-ray emission from X-ray binaries in the soft state \citep[see e.g.,][for a description of soft state variability in X-ray binaries]{Uttley05, Uttley11}. A natural explanation for this \citep[put forward in][]{MummeryBalbus22} is that this is a result of the X-ray flux of an evolving TDE disk being observed in  the Wien tail of the spectrum (i.e., it is observed at energies $E$ larger than it's peak temperature $kT_p$), meaning that the observed flux depends exponentially on the temperature of the hottest disk regions (whereas an observation of an X-ray binary in the soft state sees the full $T^4$ bolometric luminosity, with no exponential factor).   Underlying disk turbulence will result in fluctuations in this maximum temperature about its mean, which will therefore be exponentially enhanced into much larger X-ray fluctuations, and will therefore be more easily observable.  

This temperature variability, as well as being strongly observationally motivated, can be seen clearly in detailed (General Relativistic Magneto-hydrodynamical, or GRMHD) simulations of the accretion process \citep[e.g.,][among many others]{Shafee08,Noble11,Schnittman16, White16, Stone20, Liska21}. To highlight the complexity of the temperature structure in the inner regions of accretion disks (the regions which produce the X-ray flux in TDEs) we reproduce a snapshot of the GRMHD simulations of \cite{MummeryStone24}, showing the temperature of the accreting gas on a logarithmic scale (in code units) in the equatorial (left panel) and vertical (i.e., fixed azimuth; right panel) planes. Clearly, on short timescales, the temperature distribution of a realistic accretion flow is not well described by a azimuthally symmetric simple power law profile, and this fact will introduce variability into observations. Indeed, as the orbital timescale of the inner regions of a disk around supermassive black hole with mass $M_\bullet \sim 10^6 M_\odot$ is $\sim$ hours, this variability is likely to be pronounced, as one is really ``seeing'' a complex two-dimensional structure with each X-ray observation.

\begin{figure}
    \centering    
    \includegraphics[width=0.48\linewidth]{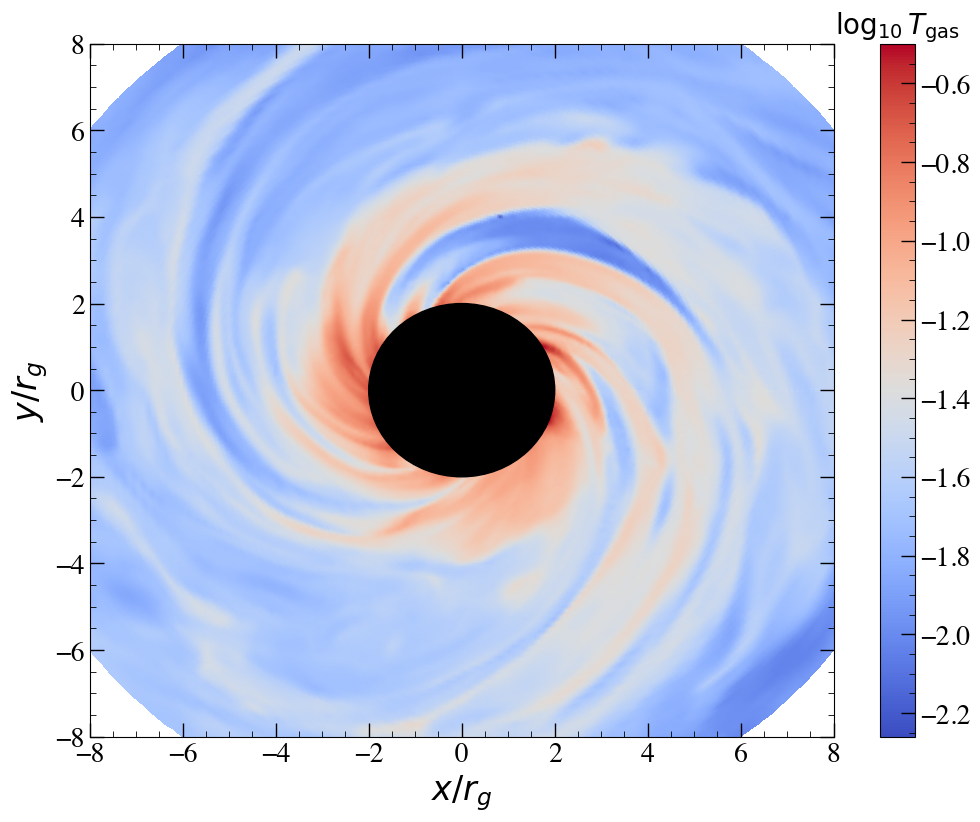}
    \includegraphics[width=0.48\linewidth]{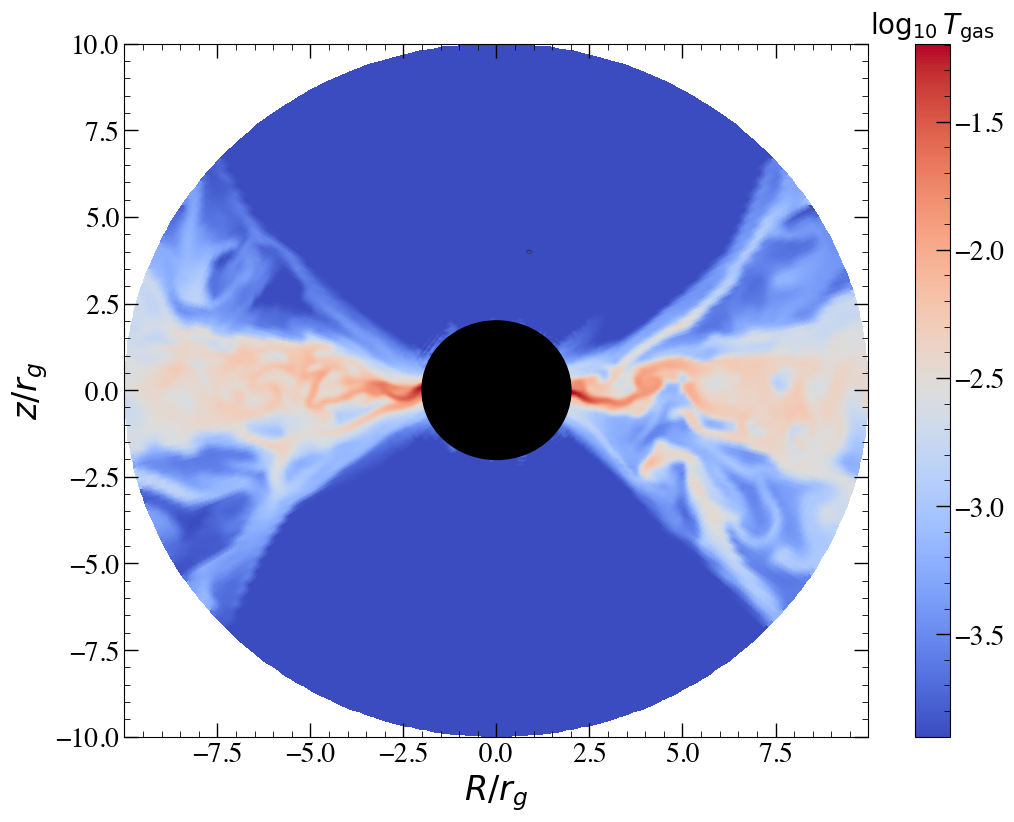}
    \caption{Realistic accretion flows are turbulent, owing to the magnetorotational instability. This results in the temperature of these flows being highly stochastic on short ($\sim$ the orbital) timescales. As this orbital timescale is typical $\sim$ hours for a TDE, this will result in stochastic, multi-timescale, variability in the observed X-ray luminosity of these sources. The panels here show two slices through a realistic GRMHD simulation of an accretion flow, with data taken from \citealt{MummeryStone24}. The temperature of the accretion flow is displayed in code units on a logarithmic scale. Clearly, capturing this variability in TDE observations will provide important constraints on physical models of the  turbulence which drives the accretion process.  }
    \label{fig:grmhd}
\end{figure}
 
In an attempt to theoretically model this variability, \cite{MummeryBalbus22} derived the probability density function that the X-ray luminosity (i.e., as observed in the Wien tail) would follow assuming that the disk temperature is log-normally distributed \citep[as required by observations][]{Uttley01, Uttley05}. This probability density function reproduced the variability of two-dimensional hydrodynamical simulations of the accretion process \citep{Turner23}, finding good agreement \citep{MummeryTurner24}. With this probability density function known from theory, and calibrated against simulations, we can ask a series of important questions regarding observations of tidal disruption systems. The first regards the distribution of the extremes of tidal disruption event light curves: how likely is it to observe a profound brightening, or dimming, episode in a list of $N$ observations of a TDE? This is, firstly, of fundamental observational interest, but also acts as an important control against which ``interesting'' variability in real systems should be contrasted with, and may even help with the interpretation of soft X-ray flares more generally (i.e., determining whether something is a TDE or not). 

The second question regards the variability structure of TDE disks over a series of different timescales. Tidal disruption events offer a wonderful opportunity to study accretion disk variability, precisely because their natural evolutionary timescales span particularly interesting observational windows. The orbital timescales in their innermost regions are of order $\sim{\cal O}(1)$ hours, which sets the absolute lower limit over which variability may be expected to be observed, and is exactly the timescale over which a typical X-ray observation is taken. The global evolution of these systems takes place over $\sim {\cal O}($months--years$)$, again a very observationally convenient timescale to probe fundamental questions of disk physics. Important questions here include:  what is the {\it shortest} timescale upon which  TDE disks vary in the X-ray? How do variability structures correlate with (e.g.) luminosity, or spectral state? 

The purpose of this  paper is to answer the first question, and to develop the framework through which the second can be answered. The statistics of light curve outliers can be analyzed using the well established theory of extreme value statistics, the results of which provide interesting, and testable, predictions for the TDE population. The second question requires a theory of stochastic light curves to be developed, which  incorporates both short-timescale correlations in X-ray observations, and allows for long-timescale secular evolution of the mean disk properties. This can be done, using a standard statistical technique known as the Cholesky decomposition method. This stochastic light curve theory is the main result of this paper. 

In the latter sections of this paper we contrast the long timescale evolution of two tidal disruption events, namely ASASSN-14li \citep{Miller15, Holoien16b} and AT2022lri \citep{Yuhan24}, with this stochastic light curve model, finding good agreement for two systems which show rather different variability properties. 

The layout of the rest of this paper is as follows. In section \ref{background} we recap the main results of \cite{MummeryBalbus22} and \cite{MummeryTurner24}, which introduces the general framework within which we shall be working. In section \ref{evs} we discuss the properties of outliers in TDE X-ray light curves, using the theory of extreme value statistics. In section \ref{lcm} we develop the a stochastic X-ray light curve model for tidal disruption events, which includes short timescale correlations and global long-term evolution in the disk properties. In section \ref{obs} we compare this theory to a pair of tidal disruption events with different variability properties, before concluding in section \ref{concs}. Some technical details are presented in Appendices \ref{CDF} and \ref{limproof}. 

\section{Mathematical background}\label{background}
The X-ray luminosity of a tidal disruption event is observed in the Wien tail, a result of the typical observed temperature $k T_{\rm obs} \approx 50-150$ eV being lower than the lower bandpass energy of an X-ray telescope $E_l = 300$ eV \citep[see e.g.,][for measurements of TDE disk temperatures]{Mummery_Wevers_23, Guolo24}. In the limit where the observed emission is in the Wien-tail of the spectrum, one does not observe a luminosity $L \propto R^2 T^4$, from which it would be simple to compute the probability density function of the luminosity. Instead, it can be shown \citep{MumBalb20a} that the luminosity follows 
\begin{equation}\label{MB}
L  = \frac{16 \pi^2  \chi_1 E_l^4}{c^2 h^ 3 f_{\rm col}^4} R_p^2 \cos i \left(\frac{kf_{\rm col} {T}_p}{E_l} \right)^\eta \exp\left(- \frac{E_l}{k f_{\rm col}{T}_p} \right) .
\end{equation} 
Here we have defined $T_p$ the hottest {\it physical} temperature in the accretion disk, and it is this temperature which is a random variable in a real disk (owing to turbulence). This physical temperature contrasts with the ``observed'' temperature $T_{\rm obs}$ by a factor $f_{\rm col}$, which is included to model radiative transfer effects and is typically $f_{\rm col} \sim 2$ for TDEs.  The radius $R_p$ corresponds to the radial location of the peak temperature, the angle $i$ is the inclination between the disk axis at the observers line of sight, while the constant $\eta$ depends on the inclination angle of the disk and the disk's inner boundary condition, and is limited to the range $3/2 \leq \eta \leq  5/2$. The constant $\chi_1 \simeq 2.19$ for classical thin disk theory, although this assumes an axisymmetric one-dimensional disk temperature profile (something which is unlikely to be true in a real turbulent flow). 

Given that we know that the disk temperature will be log-normally distributed \citep[or at least believe that we know this, with support from both observations][and first principle simulations \citealt{Turner23}]{Uttley05}, the mathematical question is then, if the high energy luminosity is described by a (dimensionless) function of the form 
\begin{equation}\label{dimX}
Y = X^\eta \, \exp(-1/X) ,
\end{equation}
and $X$ is a random variable, how is $Y$ distributed? 
In the above expression $Y = L/L_0$ is a normalised luminosity,  $X = kf_{\rm col} T_p/E_l$ a normalised temperature, and  $L_0$ is a  disk temperature independent constant defined by
\begin{equation}\label{constdef}
L_0 \equiv {16\pi^2 \chi_1 \over c^2 h^3 f_{\rm col}^4 } E_l^4 R_p^2 \cos i.  
\end{equation}
This question can be answered exactly as we know (or believe we know) how the disk temperature is distributed. It is a log-normal variable. In other words the dimensionless temperature is distributed according to
\begin{equation}
    p_X(x; \mu_N, \sigma_N) = {1\over \sqrt{2\pi \sigma_N^2 x^2}} \exp\left(-{(\ln x - \mu_N)^2\over 2\sigma_N^2} \right) . 
\end{equation}
The answer to this question was found by \cite{MummeryBalbus22}, and the X-ray luminosity is distributed according to 
\begin{equation}\label{xray_dist}
p_Y(y; \eta, \mu_N, \sigma_N) = {\eta^{-1} y^{-1} \over \sqrt{2\pi} \sigma_N \left(1 + W(z) \right) } 
\exp\left[- \left(W(z) + \eta^{-1}\ln(y) - \mu_N\right)^2 \Big/ 2\sigma_N^2\right] ,
\end{equation}
where $W(z)$ is the Lambert W function \citep{Corless96}, and $z = \eta^{-1} y^{
-1/\eta}$. The parameters $\mu_N$ and $\sigma_N$ are mathematically convenient to define, but not explicitly observable. They are related to the more readily observable parameters $\mu_T$ (the mean of the peak disk temperature distribution) and $\sigma_T^2$ (the variance in the peak disk temperature distribution), by 
\begin{align}
    \mu_N &= \ln\left({k_B\mu_T \over E_l }{\mu_T \over \sqrt{\mu_T^2 + \sigma_T^2}} \right), \\ 
    \sigma_N^2 &= \ln\left(1 + { \sigma_T^2 \over \mu_T^2}\right). 
\end{align}
This prediction of thin disk theory has been verified by comparing it to first principles two-dimensional simulations \citep{Turner23} of accretion flows \citep{MummeryTurner24}. It is this base probability density function from which we can build the distribution functions of the outliers of TDE lightcurves, and can construct a light curve model which incorporates stochastic variability and which is valid on short and long timescales.

\section{Light curve outliers and the extreme value statistics of TDEs}\label{evs}
The probability density function $p_L(\ell)$ defined above can be shown to have a large ``excess kurtosis'' \citep{MummeryBalbus22}, which means that it shows a high propensity for outliers (when compared to a normal distribution). In a physical system described by $p_L(\ell)$ what this means is that high amplitude variability will be observed regularly.

Given a stochastic light curve of a tidal disruption event, with each measurement drawn from some underlying probability density function $p_L(\ell)$, a natural question to ask concerns the statistical properties of its outliers, namely the maximum and minimum observed luminosities for a given set of $N$ observations. The mathematical framework for answering such questions is known as {\it extreme value statistics}, and we discuss its implications here. 

For a given probability density function $p_L(\ell)$, the probability that an observation $L$ is less than $\ell$ is given by the cumulative distribution function of $L$, or explicitly 
\begin{equation}
    {\rm Pr}[L \leq \ell] = \Phi_L(\ell) \equiv \int_0^\ell p_L(\ell')\, {\rm d}\ell'. 
\end{equation}
An explicit form for the cumulative distribution function (and its inverse) for the TDE X-ray luminosity distribution is given in Appendix \ref{CDF}. Therefore, in a sequence of $N$ observations, the probability that $\ell$ is the maximum luminosity observed is equal to the probability that each of the observations $\{L_1, L_2, \dots, L_N\}$ is less than $\ell$. Explicitly therefore 
\begin{equation}
    {\rm Pr}[\ell = {\rm max}] = {\rm Pr}[L_1, L_2, \dots , L_N \leq \ell] = {\rm Pr}[L_1\leq\ell]{\rm Pr}[L_2\leq\ell]\dots{\rm Pr}[L_N\leq\ell] = ({\rm Pr}[L\leq\ell])^N ,
\end{equation}
where in the final equality we have assumed that each measurement is independent (correlated observations will be discussed in a following section). The probability density function of the {\it maximum} of a set of $N$ observations is therefore given (by definition) by
\begin{equation}
    p_{L_{\rm max}}(\ell_{\rm max} | N) = {\partial \over \partial \ell} \Big({\rm Pr}[\ell = {\rm max}]\Big) = {\partial \over \partial \ell} \Big[({\rm Pr}[L\leq\ell])^N\Big],  
\end{equation}
or explicitly 
\begin{equation}
    p_{L_{\rm max}}(\ell_{\rm max} | N) = N \, (\Phi_L(\ell_{\rm max}))^{N-1} \, p_L(\ell_{\rm max}) .
\end{equation}
An analogous statement can be made about the minimum of the observed luminosities, $\ell_{\rm min}$. The probability that a given observation $L$ is greater than $\ell$ is 
\begin{equation}
    {\rm Pr}[L \geq \ell] = 1 - {\rm Pr}[L\leq \ell] = 1-  \Phi_L(\ell). 
\end{equation}
The probability that in a given set of $N$ observations that the luminosity $\ell$ is the {\it minimum} observed can be calculated following identical reasoning as above, leading to a probability density function of the minimum luminosity given by
\begin{equation}
    p_{L_{\rm min}}(\ell_{\rm min}| N) =  N \, (1 - \Phi_L(\ell_{\rm min}))^{N-1} \, p_L(\ell_{\rm min}) .
\end{equation}

\begin{figure}
    \centering
    \includegraphics[width=0.48\linewidth]{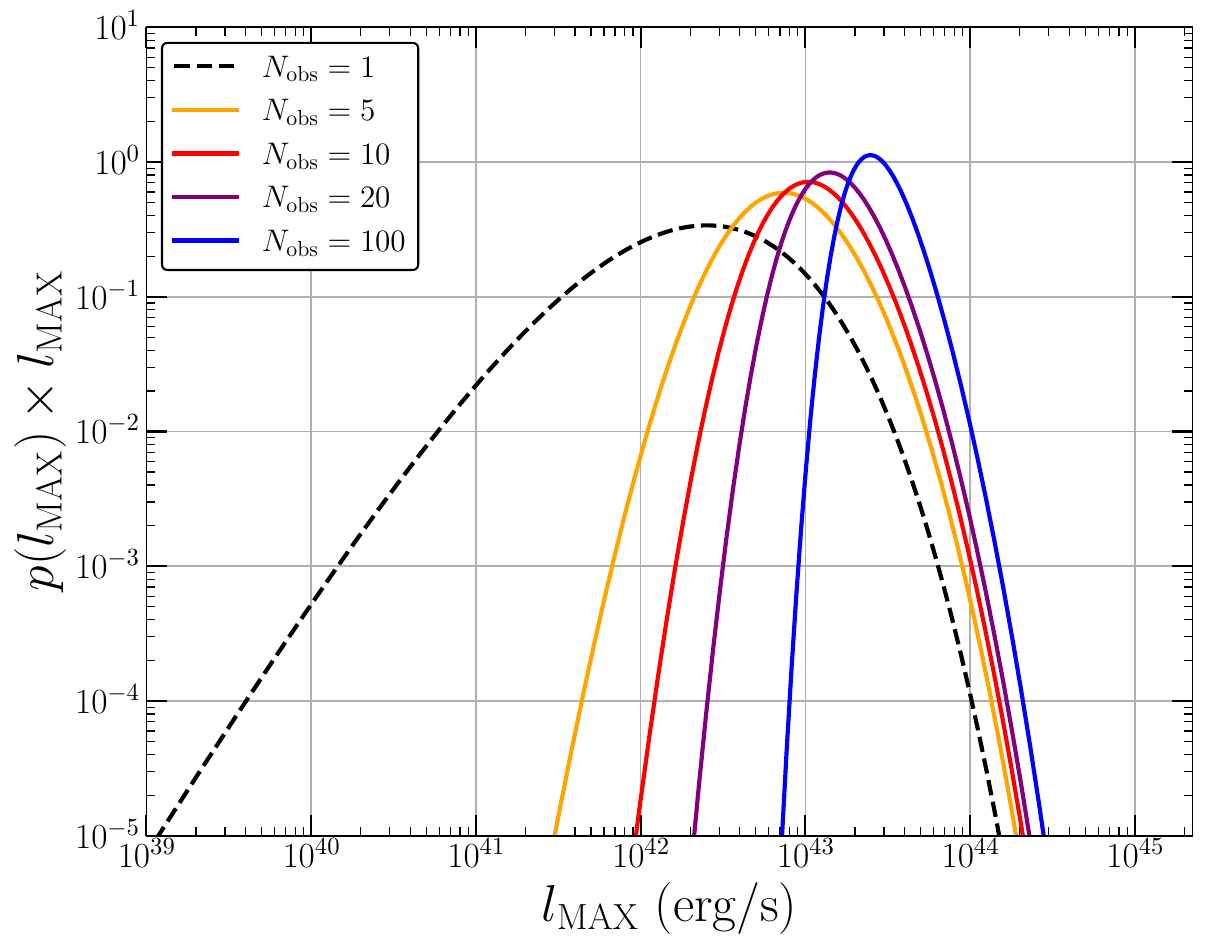}
    \includegraphics[width=0.48\linewidth]{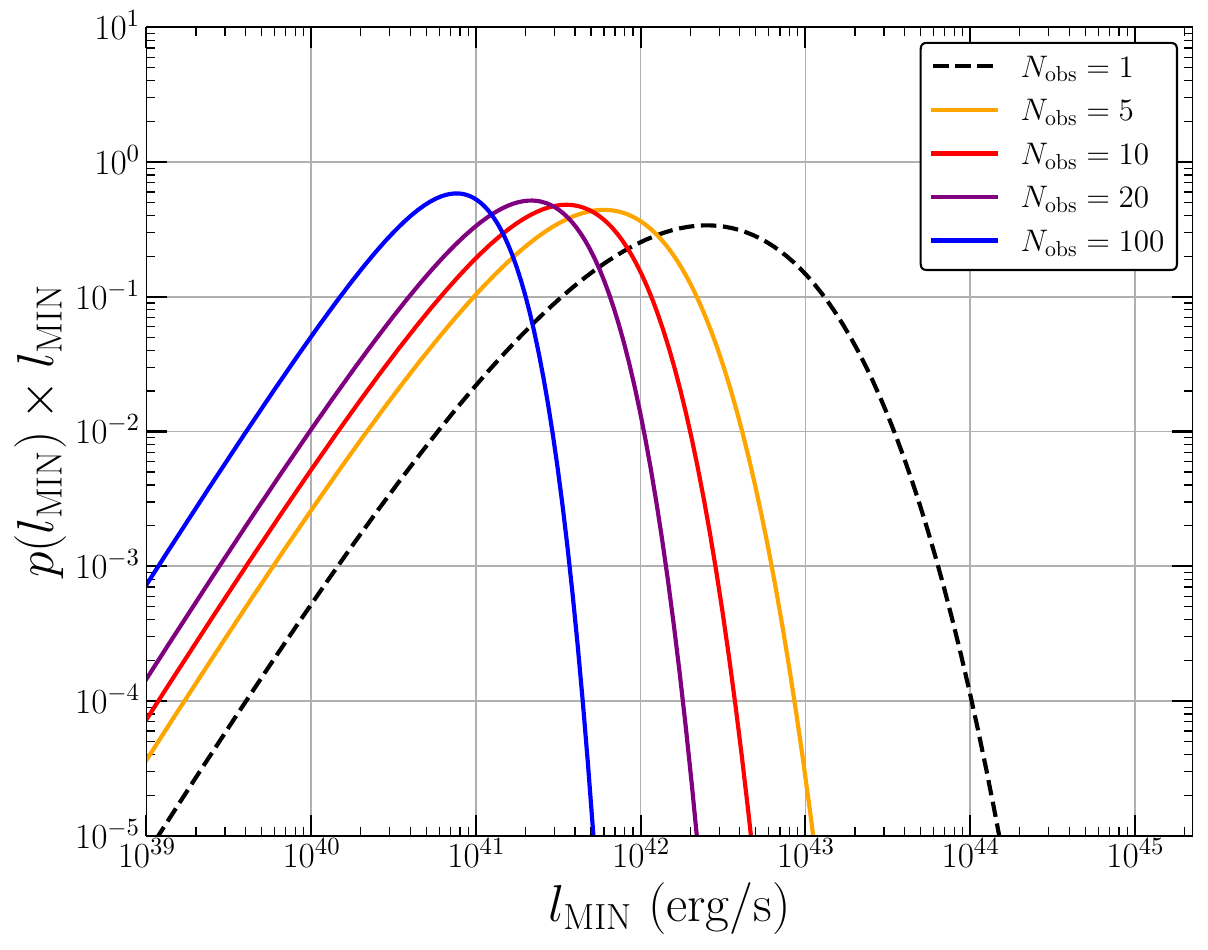}
    \caption{The distributions of the outliers of a series of $N$ observations of an accretion disk with temperature $k\mu_T = 90$ eV, with temperature variance $k\sigma_T = 20$ eV. The black hole mass was taken to be $M_\bullet = 10^6 M_\odot$.  The left hand panel shows the distribution of the {\it brightest} outliers, while the right hand panel shows the distribution of the {\it dimmest} outliers. The black dashed curve on both panels is for $N = 1$ observations and, as it must be, is identical for both. Note that the distribution of the bright outliers saturates around $L_{X, \, {\rm max}} \sim 10^{44}$ erg/s, while the distribution of dimming episodes tends to zero without a saturating bound. This is a natural result of the shape of a typical disk spectrum and can be simply understood (see text).  It is important to stress that this analysis assumes that each of the $N$ observations are uncorrelated with one another.  }
    \label{fig:evs}
\end{figure}

The properties of these outlier distributions highlight an interesting point of physics, which can be clearly seen in Figure \ref{fig:evs}, where we plot the distributions of the outliers of a series of $N$ observations of an accretion disk with temperature $k\mu_T = 90$ eV, with temperature variance $k\sigma_T = 20$ eV. The black hole mass was taken to be $M_\bullet = 10^6 M_\odot$.  Namely, the most interesting result is that while the distribution of the bright outliers saturates around $L_{X, \, {\rm max}} \sim 10^{44}$ erg/s, the distribution of dimming episodes tends to zero, without any saturating bound, in the limit of $N \gg 1$.  What this means observationally is that large amplitude {\it dimming} events should be common, and these merely represent natural variability in a turbulent flow, coupled with the particular effects of observing a system in the Wien tail. This property of Wien tail variability was first discussed in \cite{MummeryBalbus22}, but it is worth revisiting the physics of it here. 

Consider the luminosity observed in the Wien tail (e.g., equation \ref{MB}), where we take $\eta = 2$ for a face on disk (the value of $\eta$ is irrelevant for the argument)
\begin{equation}
    L \propto T^2 \exp(-E_l/kT), 
\end{equation}
and consider a large {\it positive} turbulent perturbation to the disk temperature $T \to T + |\delta T|$ (where we use the absolute value $|\delta T|$ to be clear of the signs of quantities involved).  Then the fractional perturbation to the luminosity is just 
\begin{equation}\label{dlp}
    \delta L_+ \equiv  {L_+ \over L} = \left({T + |\delta T|\over T}\right)^2 \exp\left( {|\delta T|\over T} {E_l \over k(T+|\delta T|)} \right) .
\end{equation}
We can isolate the important scaling by considering  the (likely unphysical) limit of $|\delta T| \gg T$
\begin{equation}
    \delta L_+ \sim \left({\delta T\over T}\right)^2 , 
\end{equation}
i.e., there is only a power-law increase in the Wien-tail luminosity for a large positive temperature perturbation, even for an extreme temperature change, as the exponential $\delta T$ factors cancel. For a large {\it negative} perturbation in the temperature $T \to T - |\delta T|$, then 
\begin{equation}
    \delta L_- \equiv  {L_- \over L} = \left({T - |\delta T|\over T}\right)^2 \exp\left(- {|\delta T|\over T} {E_l \over k(T-|\delta T|)} \right), 
\end{equation}
or again by taking an extreme limit (while of course restricting to $|\delta T|\ll T$) and isolating the key dependence 
\begin{equation}
    \delta L_- \sim  \exp\left(- {|\delta T| \over T} {E_l \over kT} \right), 
\end{equation}
i.e., an {\it exponential} suppression of the luminosity. This exponential-suppression and power-law-growth is the physical origin of the differing outlier distributions shown in Figure \ref{fig:evs}, and more broadly this results in {\it asymmetric} X-ray luminosity distributions. 

We note two observational facts at this stage of the analysis. The first is that the luminosity scale $L_{X, \, {\rm max}} \sim 10^{44}$ erg/s, where the positive temperature outliers saturate, is precisely the luminosity at which there is a break in the TDE X-ray luminosity function \citep{Guolo24, Grotova25}, with TDEs with X-ray luminosities $L_X = 10^{45}$ erg/s 100 times intrinsically rarer than those with X-ray luminosities $L_X = 10^{44}$ erg/s. This result was first predicted in \cite{Mum21}. The second is that short timescale large amplitude dimming events have recently been detected in a TDE by \cite{Yuhan24}, which is supportive of the framework put forward here. To determine the properties of outliers on short timescales, specifically on timescales upon which the disk evolution may be correlated, a more detailed theory must be developed however, which is the purpose of the next section. 

\section{Stochastic light curves with temporal correlations }\label{lcm}
\subsection{Stochastic evolution (short timescales)}
Given that we have a probability density function for the luminosity, the most naive approach to computing a single realization of a light curve (by which we mean a list $\vec \ell = \{\ell_1, \ell_2, \dots, \ell_N\}$ of $N$ observations of the luminosity at times $\vec t = \{t_1, t_2, \dots, t_N\}$) would be to sample a set of $N$ uniformly distributed random numbers between 0 and 1, $u_i \sim U(0, 1)$, and then use the inverse cumulative distribution function of the luminosity distribution $l_i = Q_L(u_i)$ to turn each value into a luminosity (see Appendix \ref{CDF} for the explicit form of the luminosity $Q$-function). However, this would treat each observation of the system as an {\it independent random variable}, something which will not be true if the system is observed on {\it sufficiently short timescales}. 

How short is sufficiently short? For typical tidal disruption events it seems likely that {\it during} a typical detailed observation with, e.g., the XMM instrument, any assumption of statistical independence would be a poor choice, but it is not clear if day-to-day variance may well be reasonably approximated this way.   The reason that this assumption of statistical independence would be a poor approximation for intra-hour variability is fundamentally physical. In an accretion flow there is a minimum timescale over which variables can be reasonably expected to vary, namely the orbital timescale of the innermost regions 
\begin{equation}
    t_{\rm orb} = {2\pi \over \Omega} = 2\pi \sqrt{r_{\rm orb}^3\over GM_\bullet} \approx 1\, {\rm hour} \, \left({r_{\rm orb} \over 10r_g}\right)^{3/2} \, \left({M_\bullet \over 4 \times 10^6 M_\odot}\right) ,
\end{equation}
where $r_g \equiv GM_\bullet/c^2$. This timescale represents an absolute minimum as the orbital time is the shortest natural timescale in an accretion flow. In reality it is likely that it would take many orbital timescales for the structure of the flow to vary substantially as, for example, the thermal timescale in a classical disk system is expected to be $t_{\rm therm} \sim t_{\rm orb}/\alpha$, with $\alpha < 1$, and the viscous timescale $t_{\rm visc} \sim t_{\rm orb} / (\alpha (h/r)^2)$ is expected to be even longer. Therefore, repeated observations of the same system on $\sim $ hour timescales cannot be treated as statistically independent, as the structure of the flow has not had time to vary over less than an orbital timescale. For TDE observations in the X-ray bands, which fundamentally probe the innermost regions of accretion disks, this is an important constraint as one really is observing only a small region of the flow and is therefore sensitive to properties that evolve on $\sim$ orbital timescales. 

Mathematically correlations can be introduced into stochastic light curves in the following manner. We define $t_{\rm corr}$ as the timescale over which we expect two observations to be substantially correlated, and further define 
\begin{equation}
    \Delta t_{ij} \equiv |t_i - t_j| \geq 0, 
\end{equation}
as the (absolute value of) the time between any two observations in a given light curve. The correlation matrix $\mathbf{C}_{ij}$ we define quite generally by 
\begin{equation}
    \mathbf{C}_{ij} \equiv {\cal C}(\Delta t_{ij}/t_{\rm corr}), 
\end{equation}
where the function ${\cal C}$ could in principle be arbitrary, but must satisfy ${\cal C}(0) = 1$ (observations taken at the same time are perfectly correlated), $\partial_x{\cal C}(x) < 0$ (observations taken at any later time get progressively less correlated), and ${\cal C}(x) \to 0$ for $x \to \infty$ (for times much longer than the correlation time all observations are completely uncorrelated). In other words, two observations with zero time difference are perfectly correlated, and the degree of correlation between any two observations tends to zero monotonically at large time differences. A simple example of a commonly used correlation function is ${\cal C}(x) = \exp(-x)$. For the remainder of this paper  we also make use of this  correlation function. 

With the correlation matrix defined, we can introduce this structure into a light curve in the following manner. First, one computes the so-called Cholesky decomposition (sometimes referred to as the matrix square root) of $\mathbf{C}$, namely one finds the lower triangular matrix $\mathbf{M}$ that satisfies (where superscript $T$ is the transpose operation)
\begin{equation}
    \mathbf{M} \mathbf{M}^T = \mathbf{C},
\end{equation}
where a lower triangular matrix has the following form
\begin{equation}
    \mathbf{M} = \left[
    \begin{matrix}
    M_{1,1} & 0 & 0 & \dots  & 0 \\
    M_{2,1} & M_{2,2} & 0 & \dots  & 0 \\
    \vdots & \ddots & \ddots & \ddots  & \vdots \\
    M_{N-1, 1} & \ddots & \ddots & M_{N-1, N-1}  & 0 \\
    M_{N,1} & M_{N,2} & \dots  & M_{N , N-1}  &M_{N,N}
    \end{matrix}\right] ,
\end{equation}
i.e., every element above the leading diagonal is set to zero. As $\mathbf{C}$ is a positive definite matrix this decomposition is uniquely defined. Various publicly available algorithms exist which perform this decomposition for a given matrix $\mathbf{C}$\footnote{For example \href{https://docs.scipy.org/doc/scipy/reference/generated/scipy.linalg.cholesky.html}{https://docs.scipy.org/doc/scipy/reference/generated/scipy.linalg.cholesky.html}}. Then, starting with a vector $\vec x$ of independent random draws from a standard normal 
\begin{equation}
    \vec x \sim {\cal N}(0, 1), 
\end{equation}
one defines the vector $\vec y$ by 
\begin{equation}
    \vec y = \mathbf{M} \vec x .
\end{equation}
The vector $\vec y$ is a vector of random numbers which inherits the required correlation matrix $\mathbf{C}$. This can be readily verified by explicit computation  
\begin{equation}
    {\rm corr}(\vec y) = \mathbb{E}[\vec y \vec y^T] = \mathbb{E}[(\mathbf{M}\vec x)(\mathbf{M}\vec x)^T] = \mathbf{M}\, \mathbb{E}[\vec x\vec x^T]\, \mathbf{M}^T = \mathbf{M} \mathbf{M}^T = \mathbf{C}. 
\end{equation}
Here ${\rm corr}(\vec y)$ is the definition of the auto-correlation matrix of a vector $\vec y$, $\mathbb{E}$ is the expectation value (linear) operator, and $\mathbb{E}[\vec x_i \vec x_j] = \delta_{ij}$ for a variable $\vec x$ distributed according to the standard normal (throughout this paper $\delta_{ij}$ is the usual  Kronecker delta). Therefore, the vector $\vec u$, defined by 
\begin{equation}
    \vec u \equiv \Phi_N(\vec y) = {1\over 2} \left[1 + {\rm erf}\left(\vec y \over \sqrt{2}\right)\right], 
\end{equation}
is a vector of (non-uniform) random numbers between 0 and 1, which again maintains the required correlation matrix $\mathbf{C}$ (here $\Phi_N$ is the cumulative distribution function of the standard normal distribution, and ${\rm erf}(z)$ is the error function). This is all that is required to compute a correlated light curve, which is then given explicitly by 
\begin{equation}\label{LC}
    \vec \ell = Q_L(\Phi_N(\mathbf{M} \vec x\,)). 
\end{equation}
To summarize, given a list of observational times $\vec t=\{t_1, t_2, \dots, t_N\}$, and an assumed correlation time of the accretion flow $t_{\rm corr}$, one computes the required correlation matrix $\mathbf{C}$,  its Cholesky decomposition $\mathbf{M}$, and samples $\vec x$ from a standard normal distribution. The vector $\vec \ell$ given by equation \ref{LC} then represents one realisation of a stochastic light curve, with physical correlations maintained. 

\begin{figure}
    \centering
    \includegraphics[width=0.47\linewidth]{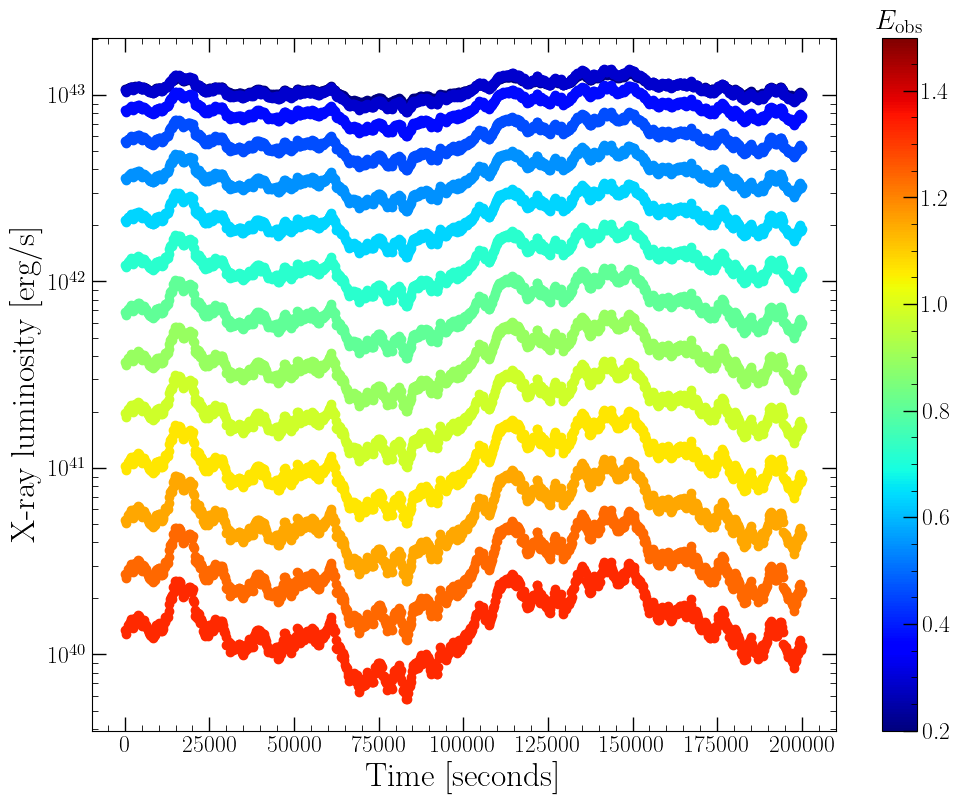}
    \includegraphics[width=0.50\linewidth]{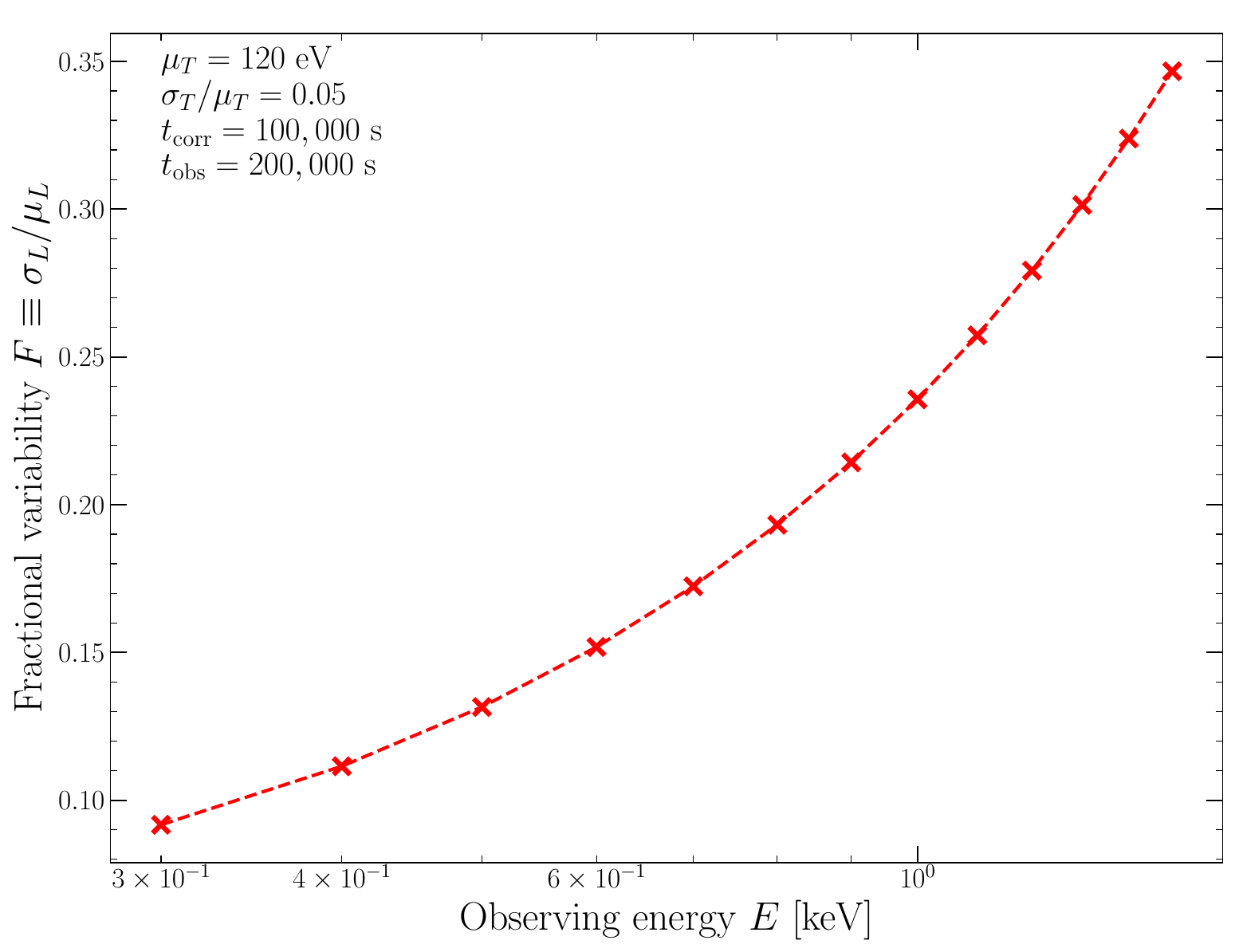}
    \caption{On the left we show a single realisation of a short-timescale X-ray light curve, with luminosity measured above a series of different observing energies (colourbar), for a disk with mean temperature $k\mu_T = 120$ eV, temperature variance of $\sigma_T/\mu_T = 0.05$ and a correlation time $t_{\rm corr} = 100,000$ seconds. The observations where taken near-continuously (sampled $N=1000$ times) over $t_{\rm obs} = 200,000$ seconds. These parameters were chosen to simulate a detailed XMM observation across different energy channels. On the right we show the fractional variability of each of the light curves, showing how the fractional variability increases with observing energy. This is a generic prediction of the theory developed here, and while the {\it values} of $F$ vary on a case-by-case basis, the trend of $F$ with energy is a robust prediction of this theory.  Both the energy dependence and short-timescale correlations are clear to see, by eye, in the left hand plot.   }
    \label{fig:light_curves}
\end{figure}

The first important computation we can perform with this model is a realisation of a short-timescale (i.e., an intra-observation) X-ray light curve, with luminosity measured above a series of different observing energies. These different energies are chosen to demonstrate the variability that would be observed across an X-ray satellite bandpass in different energy channels. We show such a realisation in Figure \ref{fig:light_curves}, for a disk with mean temperature $k\mu_T = 120$ eV, temperature variance of $\sigma_T/\mu_T = 0.05$ and a correlation time $t_{\rm corr} = 100,000$ seconds. The black hole mass was taken to be $M_\bullet = 10^6 M_\odot$. The different energies are denoted by the colourbar.  The observations where taken near-continuously (i.e., $N=1000$ points were sampled) over $t_{\rm obs} = 200,000$ seconds. We chose these values with the intention of simulating a detailed XMM observation. On the left of Figure \ref{fig:light_curves} we show the actual light curves, and on the right we show the fractional variability of each of these light curves, highlighting how the fractional variability increases with observing energy, a simple prediction of the framework put forward here.  Both the energy dependence and short-timescale correlations are clear to see, by eye, in the left hand light curve plot. 

We always find that the fractional variability $F \equiv \sigma_L/\mu_L$ increases with observing energy \citep[this is another simple result of the shape of the Wien tail, and has been proved rigorously][]{MummeryBalbus22} and this can be considered a prediction of the framework put forward here. To be explicit, the origin of this energy dependence can be simply seen in equation \ref{dlp} (reproduced here)
\begin{equation}
    \delta L =  {L(T+\delta T) \over L(T)} = \left({T + \delta T\over T}\right)^2 \exp\left( {E\over kT} {\delta T \over (T+\delta T)} \right) , 
\end{equation}
where the factor $E/kT$ in the exponential highlights that variability at higher observing energies is, at fixed mean disk temperature (a very good approximation  for data taken over a few hours),  exponentially enhanced over that at lower observing energies. 

Note that, while this appears a relatively trivial result, it is nevertheless a robust and important test of the framework put forward here. There are other ways one could imagine the X-ray light curve of a tidal disruption event varying which would not necessarily induce this energy-dependent signature. For example a global change in the size/geometry of the disk, time varying obscuration of the flow, or plausibly precession of the disk. It is explicitly a result of the assumed turbulent variability of the disk temperature that this energy dependence is induced, and therefore this energy dependence can be used as a probe of the temperature variability (if confirmed by observations). 

\begin{figure}
    \centering
    \includegraphics[width=0.48\linewidth]{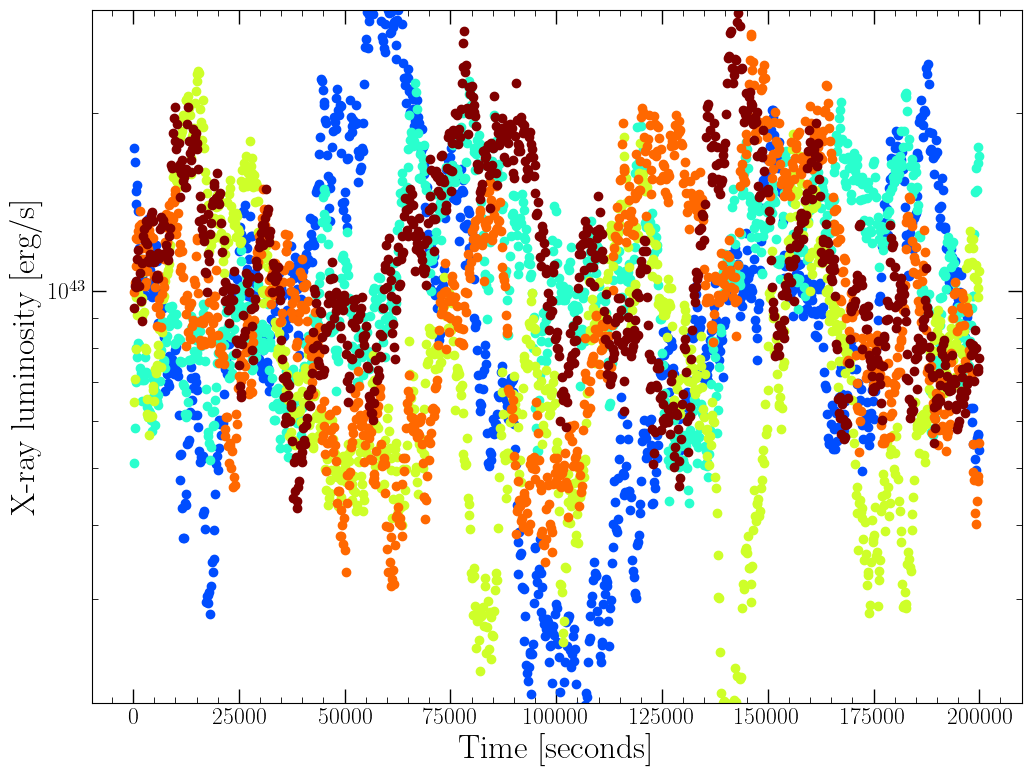}
    \includegraphics[width=0.48\linewidth]{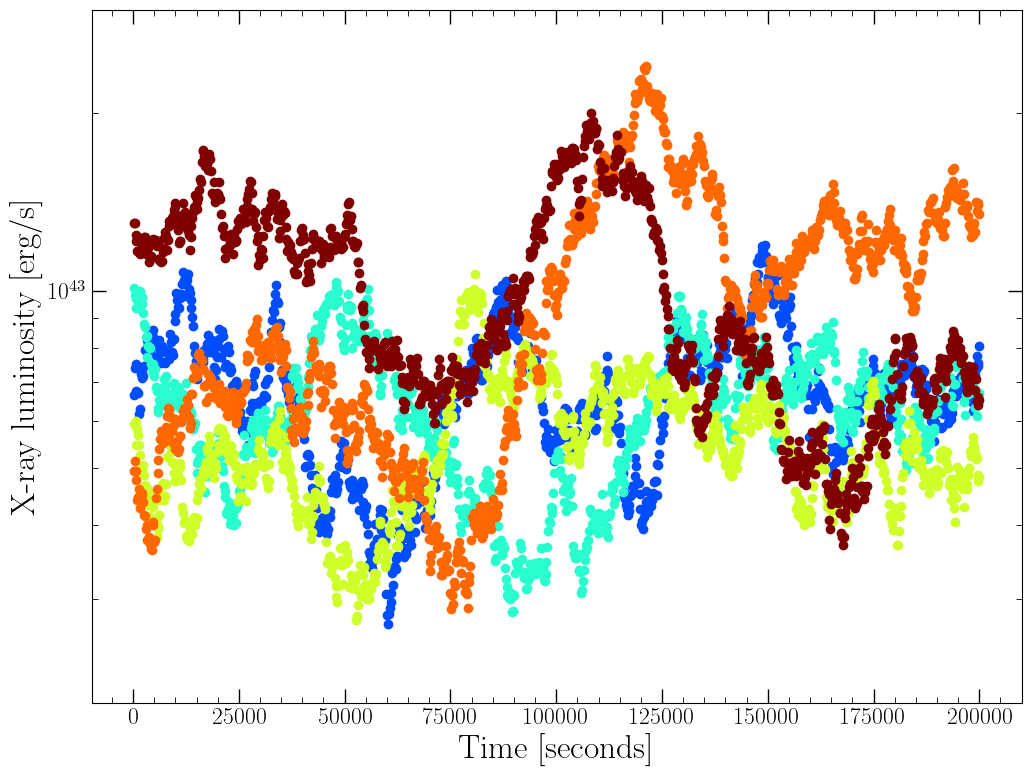}
    \includegraphics[width=0.48\linewidth]{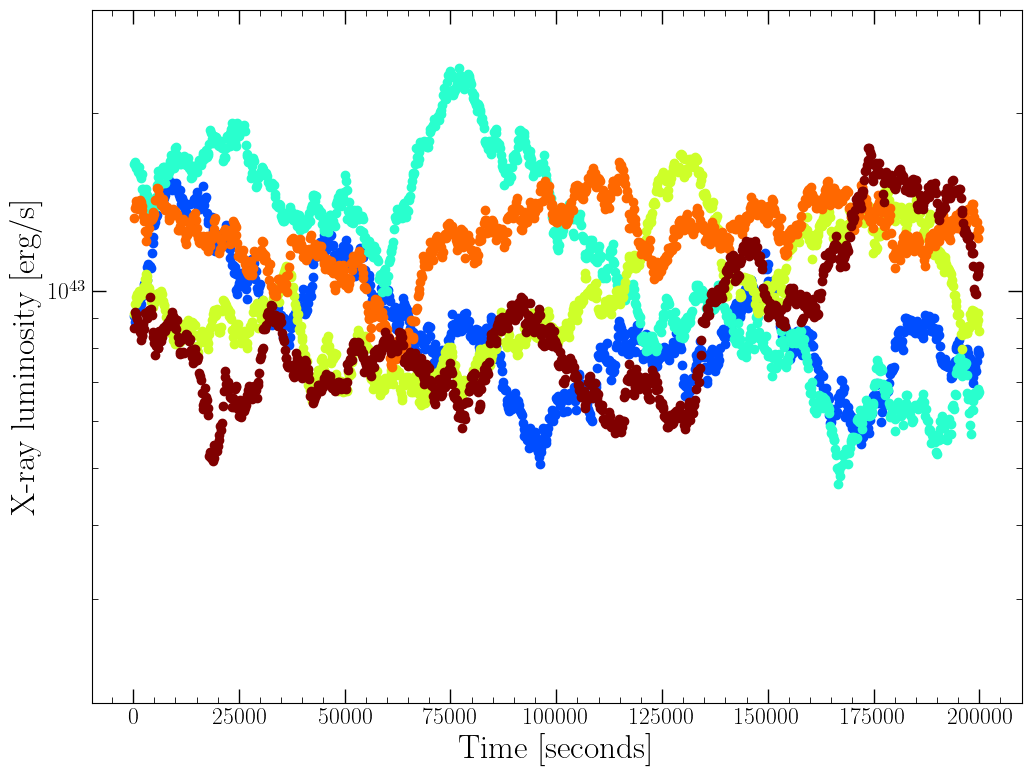}
    \includegraphics[width=0.48\linewidth]{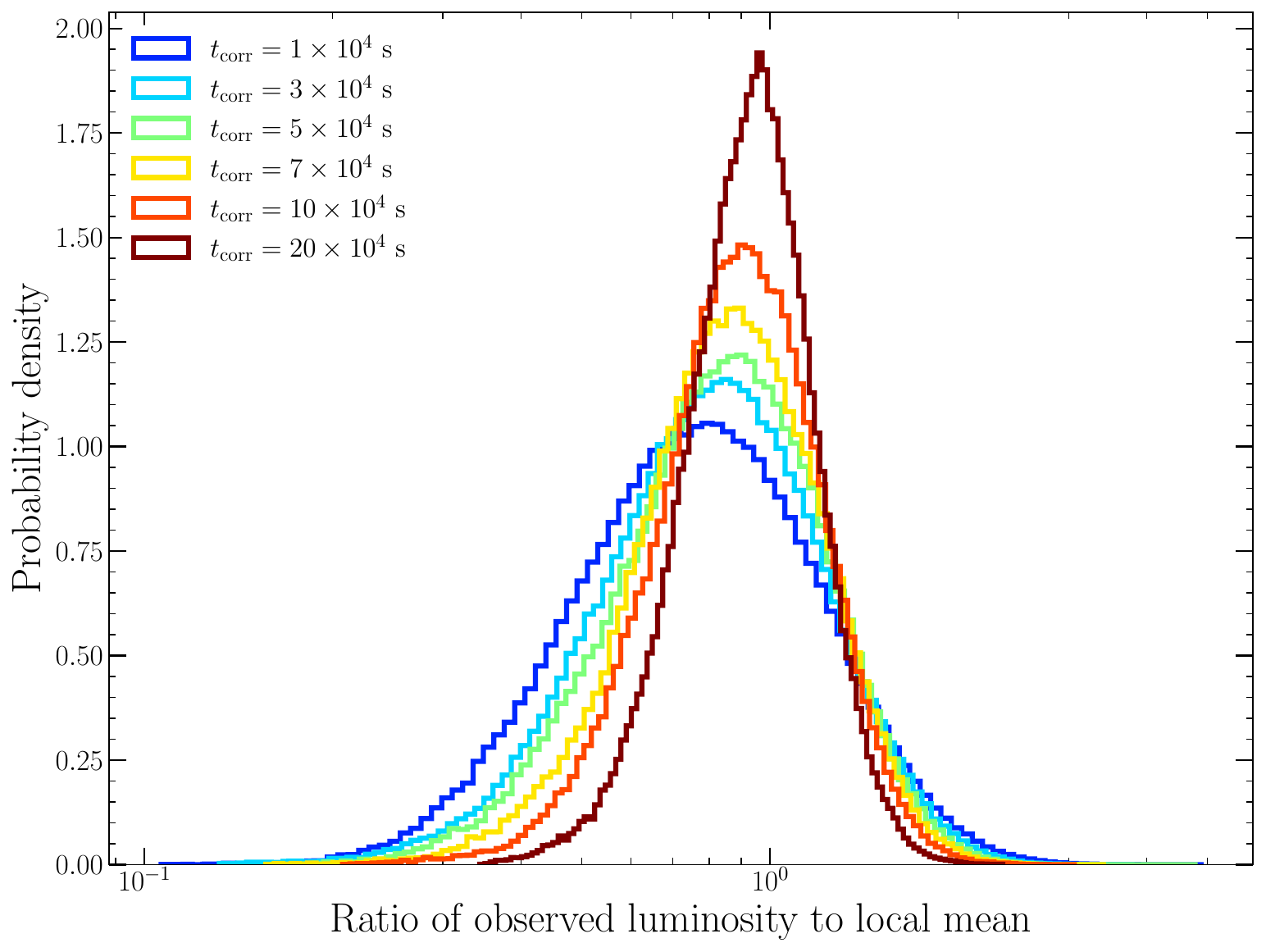}
    \caption{Five random realisations of stochastic, short-timescale, X-ray light curves for different correlation times of the flow, each for mean disk temperature $k \mu_T = 120$ eV and for fractional temperature variability $\sigma_T/\mu_T = 0.1$. The (most variable) upper left plot has $t_{\rm corr} = 10^4$ seconds, the (less variable) upper right plot has $t_{\rm corr} = 5\times 10^4$ seconds, and the (least variable) lower plot has $t_{\rm corr} = 10^5$ seconds. Note that while each set of light curves are generated with fixed fractional temperature variability, short correlation timescales produce noticeably larger variance in a given observational window. Given that the orbital timescale of the inner regions of a supermassive black hole disk can be of order $t_{\rm orb} \sim {\cal O}(10^3-10^4)$ seconds, short-timescale observations of TDEs will likely  provide an extremely interesting probe of the natural timescale of turbulent variability in accretion disks (in units of the orbital timescale), a question of fundamental interest in disk physics. Note that the vertical axes of each panel are the same.  On the lower right plot we show the distribution of the ratio of each observed luminosity to the mean across the observing window (i.e., the distribution of $R=\{L_j\}/\left\langle \{L_j\}\right\rangle$ for a list of observations $\{L_j\}$), averaged over 500 random realisations of the light curves, as a function of the assumed correlation timescale of the flow (displayed on plot). All other parameters are the same as for the three other plots.   }
    \label{fig:correlation_time}
\end{figure}

\begin{figure}
    \centering
    \includegraphics[width=0.75\linewidth]{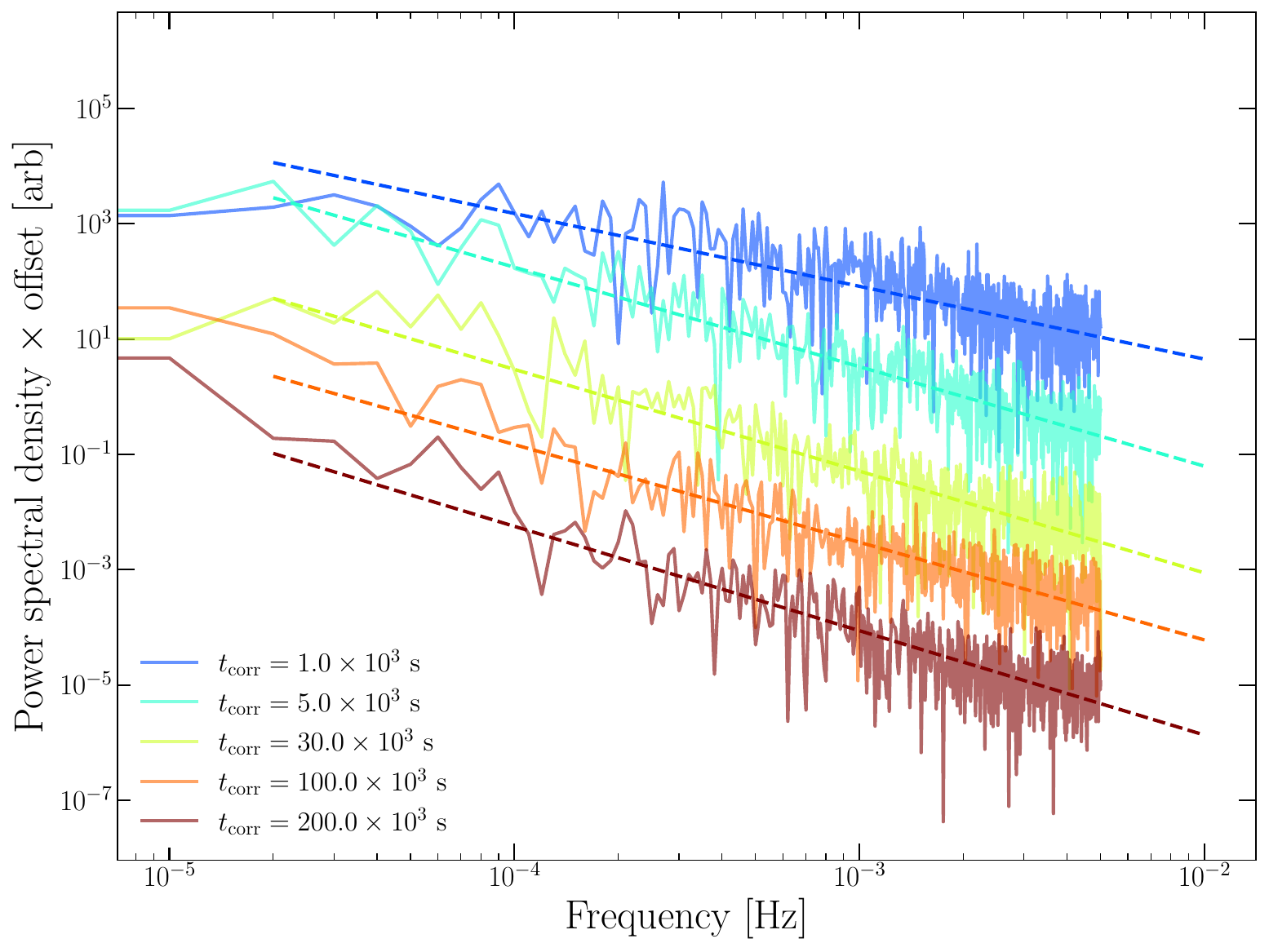}
    \caption{The power spectral density plotted against frequency of five different realisations of stochastic, short-timescale, X-ray light curves for different correlation times of the flow (denoted on plot). These light curves are generated in an identical manner to those shown in Figure \ref{fig:correlation_time}. The amplitude of the PSDs are re-normalised by a multiplicative offset for visibility purposes, while the frequency axis is in physical units. Dashed lines show simple power-law fits to the PSD of each light curve. While all light curves are formally described by ``red'' noise, shorter correlation times (relative to the typical observing cadence) produce ``whiter'' noise (see text for more discussion).     }
    \label{fig:PSD}
\end{figure}

We note here that tidal disruption events offer a potentially unique probe of the physical timescale over which accretion flows may vary (in units of the orbital timescale). This is a fundamentally important question in disk physics, as it is ultimately related to the physics of (turbulent) angular momentum transport in disks, and therefore the (micro)physics of the accretion process itself. 

To see why this is the case, consider Figure \ref{fig:correlation_time} which shows five random realisations of stochastic, short-timescale, X-ray light curves for different correlation times of the flow. Each panel is produced with both the same mean disk temperature $(k \mu_T = 120$ eV$)$ and the same fractional temperature variability $(\sigma_T/\mu_T = 0.1)$. The (most variable) upper left plot has $t_{\rm corr} = 10^4$ seconds, the (less variable) upper right plot has $t_{\rm corr} = 5\times 10^4$ seconds, and the (least variable) lower plot has $t_{\rm corr} = 10^5$ seconds. Note that the vertical axes of each panel are the same. All of these values are, {\it apriori}, perfectly plausible values for a typical TDE disk, as they represent $\sim{\cal O}$(a few -- few ten's) of orbital timescales.  While each set of light curves are generated with fixed fractional temperature variability, short correlation timescales produce noticeably larger variance in a given observational window\footnote{Mathematically this is simple to understand, as in the limit $t_{\rm corr} \to 0$ we have $\mathbf{C}_{ij}\to \delta_{ij}, \, \mathbf{M}_{ij}\to \delta_{ij}$ and therefore every point is a unique draw from a probability distribution with large intrinsic variance (e.g., see Figure \ref{fig:evs}).}. 

The effect of a changing the correlation timescale can be more quantitatively seen in the lower right plot of Figure \ref{fig:correlation_time}, where we display the distribution of the ratio of each observed luminosity to the mean across the observing window (i.e., the distribution of $R=\{L_j\}/\left\langle \{L_j\}\right\rangle$ for a list of observations $\{L_j\}$), averaged over 500 random realisations of each light curve, as a function of the assumed correlation timescale of the flow (displayed on plot). All other parameters are the same as for the three other plots. We see that in the limit $t_{\rm corr} \to 0$ there is a shift towards negative relative luminosities (as can be understood by considering the extreme value statistics discussed earlier), and a broadening of the distribution. In the opposite limit $t_{\rm corr} \to \infty$ the relative luminosities will all be equal to 1, as each observation will be perfectly correlated, meaning the distribution $p_R \to \delta(R - 1)$.  

Given that the orbital timescale of the inner regions of a supermassive black hole disk can be of order $t_{\rm orb} \sim {\cal O}(10^3-10^4)$ seconds, observations of TDEs will likely  provide an extremely interesting probe of the natural timescale of turbulent variability (in units of the orbital timescale), a question of fundamental interest in disk physics.   We believe that such an analysis is of real interest, and encourage such tests to be performed on existing TDE data sets. 

Note that there is a natural degeneracy between the amplitude of the fractional temperature variability $\sigma_T/\mu_T$ and the correlation time $t_{\rm corr}$ in setting the observed fractional variability of the X-ray emission $\sigma_X/\mu_X$ {\it within one observational window} (i.e., increasing $\sigma_T/\mu_T$ or decreasing $t_{\rm corr}$ will both increase the X-ray variability on short timescales). However, only an increase in $\sigma_T/\mu_T$ will increase the {\it long time} variability in the X-ray emission (i.e., the variability on times $t\gg t_{\rm corr}$), and so in principle if one has access to both detailed short timescale observations and a long timescale light curve, these two effects can be distinguished. 

Another common approach to quantifying the variability of time-series data is by analyzing the power spectral density (PSD) of the signal. The PSD, denoted $S(f)$ here, is derived from the Fourier transform of the time series $l(t)$, and is formally defined as the amplitude of 
\begin{equation}
    \hat l(f) \equiv \int_{-\infty}^{+\infty} l(t) \, \exp(-2\pi i ft)\, {\rm d}t,
\end{equation}
i.e., 
\begin{equation}
    S(f) \equiv \left| \hat l(f) \right|^2. 
\end{equation}
The absolute amplitude of $S$ is not of much inherent interest but the frequency-dependence of $S$ is, as it can help quantify the type of physical process producing the time-series. The most common characterization is the ``color'' of the noise, whereby a power-law is fit to the PSD $S=Af^{-n}$, and the value of $n$ is used as a diagnostic.  White noise is defined as $n = 0$, while ``redder'' noise is defined by $0 < n < 2$, with $n=1$ often referred to as ``pink'' noise. White noise is formally uncorrelated, while redder noise has an increasing degree of correlation. 

In Figure \ref{fig:PSD} we show the power spectral density plotted against frequency of five different realisations of stochastic, short-timescale, X-ray light curves for different correlation times of the flow (denoted on plot). These light curves are generated in an identical manner to those shown in Figure \ref{fig:correlation_time}. The amplitude of the PSDs are re-normalised for visibility purposes, while the frequency axis is in physical units. We observe from this plot that the shorter the correlation timescale of the disk, the ``whiter'' (i.e., shallower slope) the PSD. This is not a property of the absolute value of the correlation timescale, but is a result of the {\it ratio} of the correlation timescale to the typical observing cadence of the light curve. This is simple to understand physically, as a shorter correlation time (at fixed observational cadence) has more ``observations'' which are uncorrelated with one another, and are therefore more readily described by white noise. Longer correlation timescales (at fixed observational cadence) are more correlated, and therefore are ``redder''. Each PSD is reasonably well described by a red-noise profile, with power-law indices $n$ varying for the choice of correlation timescale, from $n \approx 1$ to $n \approx 2$ across the different realisations. If the cadence were reduced (i.e., so that frequencies $f \gtrsim 10^{-3}$ Hz were excluded) the shortest correlation timescale data would be well approximated by white noise. 

It is natural therefore to expect that X-ray light curves of TDEs will, depending on the precise details of the correlation time and typical observing cadence, be well described by simple power-law noise somewhere between the extremes of red and white profiles. 

\subsection{Secular evolution (long timescales)}
In addition to short timescale stochastic variability (which as we have discussed will likely be correlated on some natural timescale in the problem) tidal disruption events undergo long-timescale evolution, owing to the transient nature of the event itself.  Relevant for our purposes, the initial onset of accretion causes the peak temperature $T_p$ to rise, before it then decays away as the disk material is accreted and the inner disk density drops with time. 

Fortunately, this long-timescale evolution is well understood, as the gross disk evolution is described by the relativistic disk equations \citep{Balbus17}, with known analytical Greens function solutions \citep{Mummery23a}. To leading order therefore we shall treat the mean temperature (which we remind the reader is an input into the luminosity probability density function) of the evolving disk as being given by these Greens function solutions. The mean evolution of the disk temperature is then given by \citep{Mummery23a}
\begin{equation}
    \mu_T(t) \approx \widehat \mu_T \, a_n \, \left({t \over t_{\rm evol}}\right)^{-n/4} \, \exp\left(-{ t_{\rm evol} \over 4t}\right) ,
\end{equation}
where $t_{\rm evol}$ is the long-term evolutionary timescale of the disk (often called the viscous timescale), $n\approx 1.2$ is the index at which the bolometric luminosity of the system decays, and $a_n = n^{n/4} e^{1/4n} \sim {\cal O}(1)$ is a normalisation amplitude chosen so that $\widehat \mu_T$ is the peak temperature reached over the evolution of the system (which occurs at a time $t = n t_{\rm evol}$). 

The long time properties of the variance in the temperature $\sigma_T^2$ are more difficult to ascertain from first principles (and are therefore more interesting). The simplest model would be to take $\sigma_T/\mu_T = {\rm constant}$ at all times, as would be expected for a system whose bolometric luminosity follows a linear rms-flux relationship \citep{Uttley05}. However, there is reason to expect that the fractional variability in the disk temperature is a function of the disk parameters, as first principles simulations suggest that this fractional variability increases with disk aspect ratio $H/r$ \citep{Turner23}. This result is relatively simple to understand in a qualitative sense, as turbulent eddies in the flow are likely to be restricted by the scale height of the disk, and therefore thicker disks are expected to have larger amplitude turbulent fluctuations, and therefore a higher fractional variability in their temperature \citep[as has been verified numerically][]{MummeryTurner24}. 

Bearing in mind these complications, in this work we shall restrict ourselves to the simplest case of $\sigma_T/\mu_T = {\rm constant}$. Note that of course this results in the variance of the disk temperature changing in line with the secular evolution of the mean of the disk temperature. 

A final important (but relatively trivial) result, which will be of use later, is that in the limit $\sigma_T/\mu_T \to 0$ we have, for any assumed correlation timescale, that our stochastic light curve $l(t)$ will always be exactly equal to $\overline L(t)$, where 
\begin{equation}
    \overline L(t) = \frac{16 \pi^2  \chi_1 E_l^4}{c^2 h^ 3 f_{\rm col}^4} R_p^2 \cos i \left(\frac{kf_{\rm col} {\mu}_T(t)}{E_l} \right)^\eta \exp\left(- \frac{E_l}{k f_{\rm col}\mu_T(t)} \right) ,
\end{equation}
is the thin disk theory prediction for the X-ray luminosity, with $\mu_T(t)$ given by the above expression. This is because, in this limit\footnote{A proof of this is relatively simple (see Appendix \ref{limproof}), and ultimately results from the fact that the temperature variance can be related to some underlying normal distribution $p_Z(z|\mu_N, \sigma_N)$, and that the normal distribution asymptotically approaches $\delta(z-\mu_N)$ in the limit $\sigma_N\to 0$.}, 
\begin{equation}
    p_{L}(l| \sigma_T\to 0) \to \delta(l - \overline L(t)) ,
\end{equation}
and so regardless of the random sampling procedure, one always recovers the thin disk evolution equation. This limit will allow us to define a well-posed mean evolution around which relative luminosity ratios can be defined in a system with long-timescale evolution, which is something to which observations may be compared. 

\subsection{Some properties of the stochastic long-term light curves}
\begin{figure}
    \centering
    \includegraphics[width=0.48\linewidth]{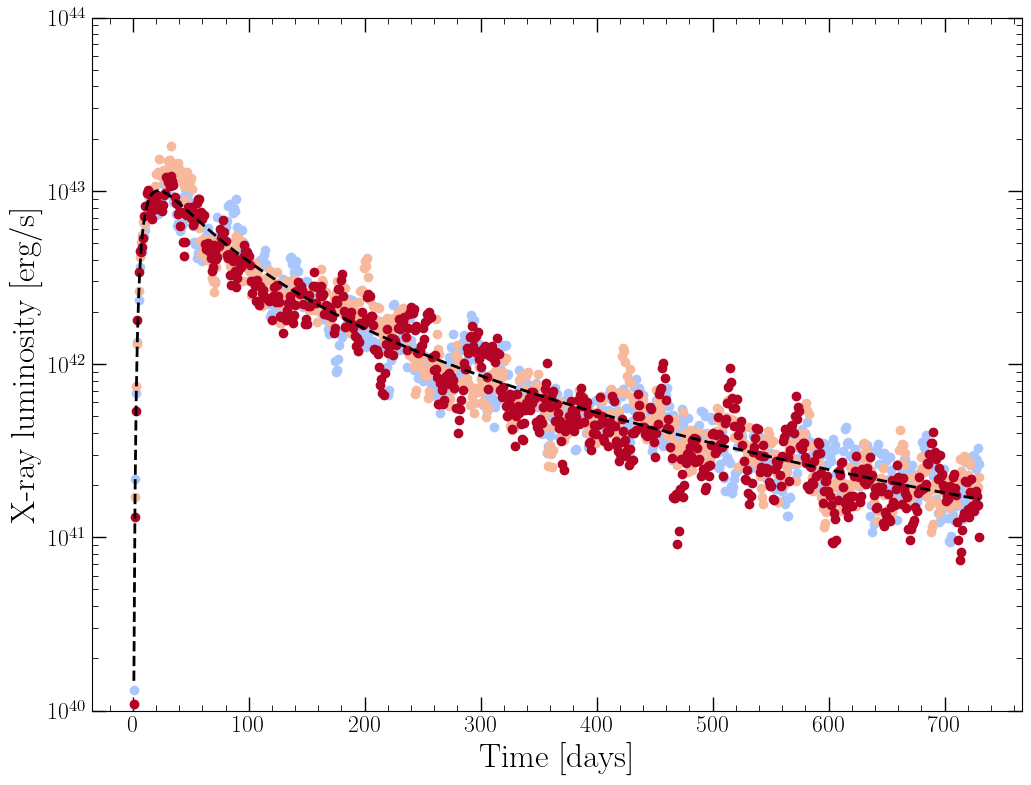}
    \includegraphics[width=0.48\linewidth]{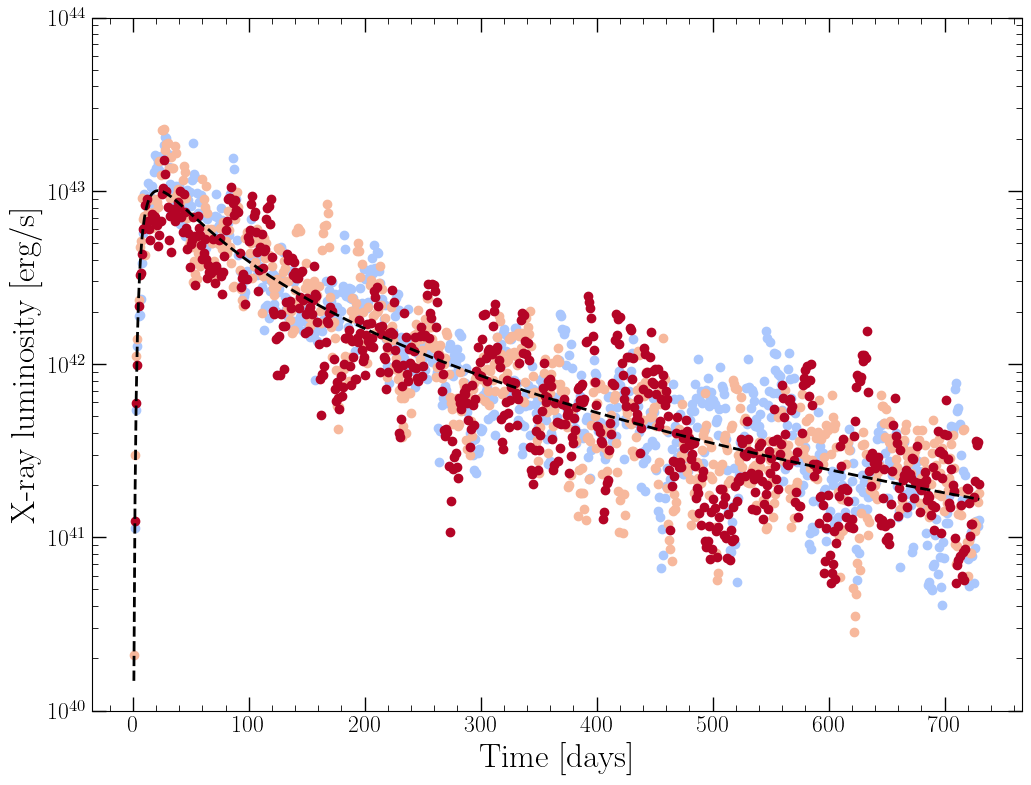}
    \includegraphics[width=0.48\linewidth]{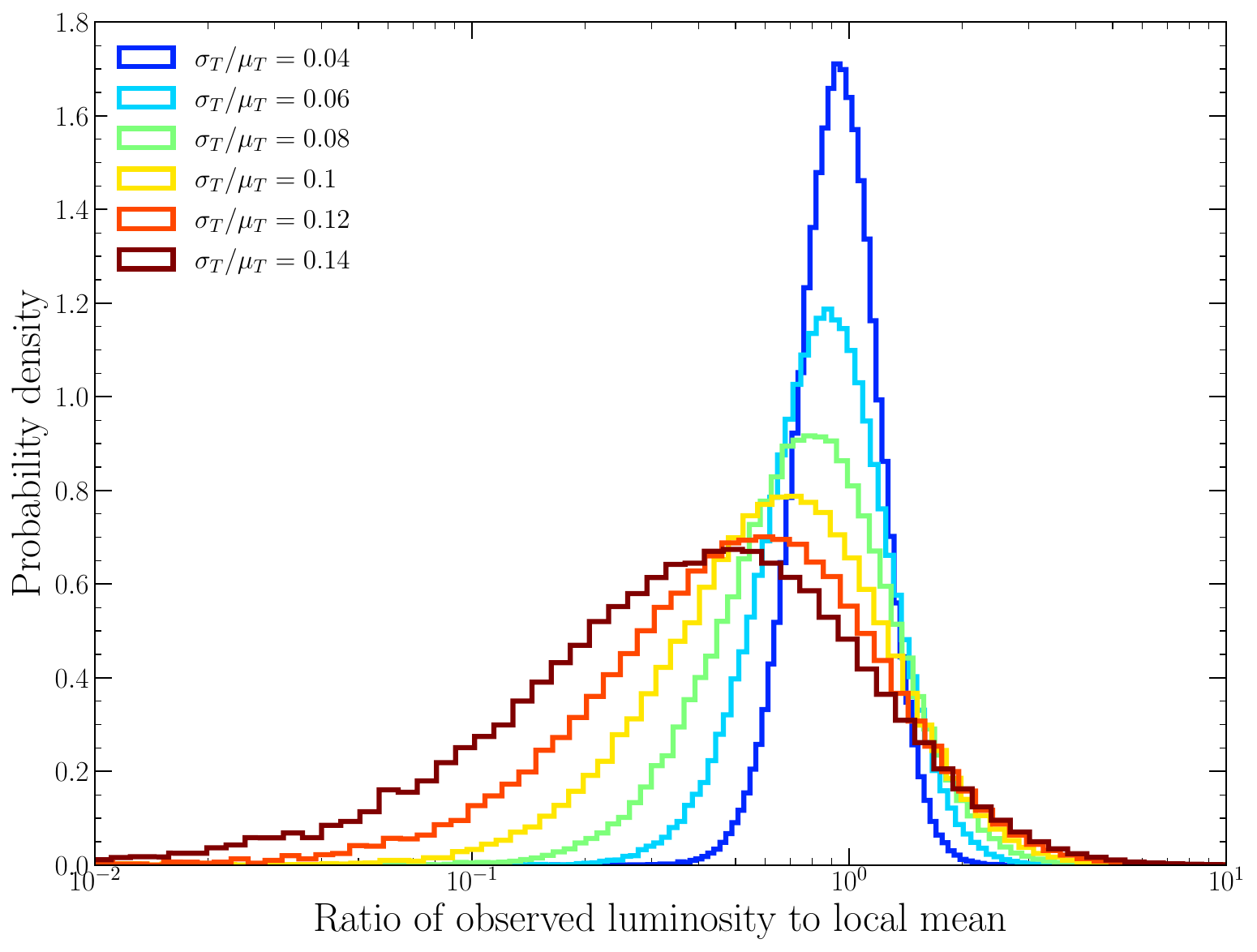}
    \includegraphics[width=0.48\linewidth]{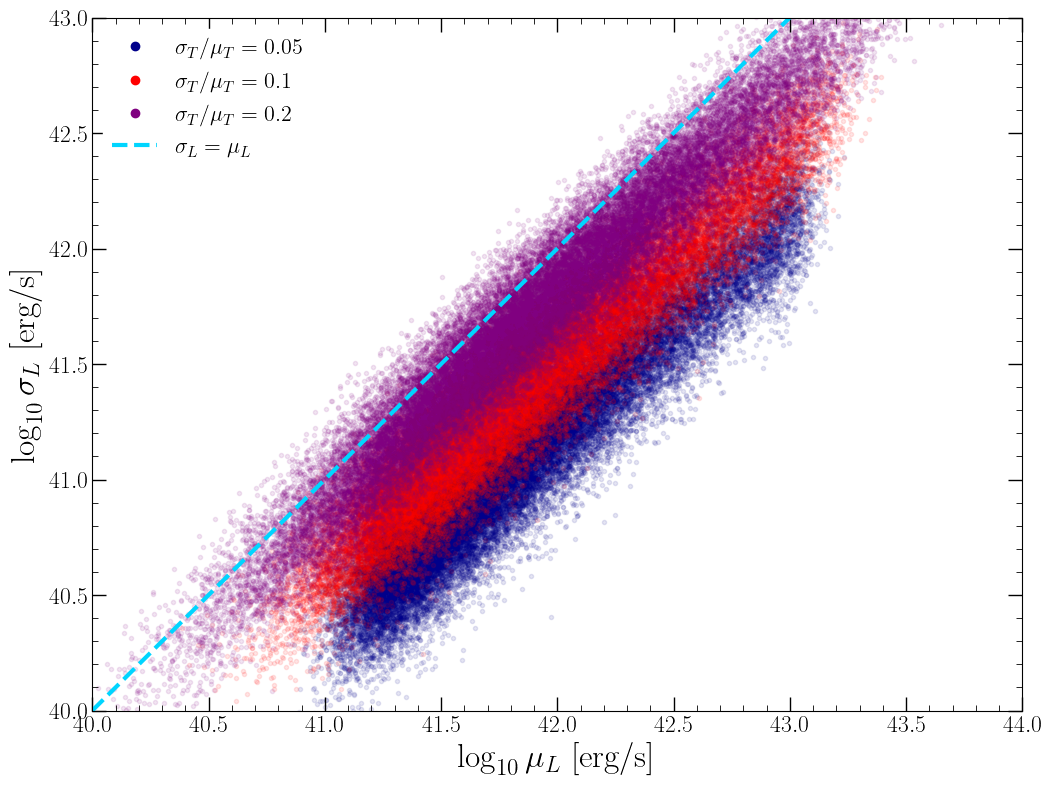}
    \caption{{\bf Upper:} Some examples of long time stochastic X-ray light curve evolution, with correlated short-timescale variability (with correlation timescale $t_{\rm corr} = 5$ days), and two different values of the fractional temperature variability $\sigma_T/\mu_T = 0.05$ (left) and $\sigma_T/\mu_T = 0.1$ (right). The peak value of the mean of the disk temperature in each evolution was taken to be $k \hat \mu_T = 120$ eV, and the evolutionary timescale was chosen so that the mean X-ray luminosity dropped by $\sim 2$ orders of magnitude over the two years, a ballpark figure taken from observations of real TDEs. Each system is assumed to be ``observed'' every day for two years. It is clear to see from this Figure that the observed X-ray luminosity variability is a sensitive function of assumed temperature variability. Note that $\sim$ order of magnitude short-timescale X-ray dimming events are seen in the right hand panel, as has also been seen in tidal disruption events. {\bf Lower:} some statistical properties of these long-timescale stochastic systems, computed using a large sample of realisations. On the left we show the distribution of the local luminosity ratio (i.e., $R=l(t)/\overline L(t)$, where $l(t)$ is a stochastic light curve realisation and $\overline L(t)$ is the mean evolution defined in the text) as a function of the assumed fractional temperature variability. On the right we show a large sample of the so-called RMS-flux relationship predicted in this theory, for three different values of the  fractional temperature variability. Note that the fractional variability in the luminosity can reach $\sigma_L/\mu_L\sim 1$ for $\sigma_T/\mu_T \sim 0.1$, a large enhancement. The parameters not explicitly varied in the lower plots are held to the same values as those in the upper plots.    }
    \label{fig:long}
\end{figure}
There are of course a very wide range of possible stochastic light curves one can generate with the various free parameters of this theory (i.e., $\Theta = [L_0, \hat\mu_T, \sigma_T, \eta, t_{\rm corr}, t_{\rm evol}, E_l]$) and we do not perform an exhaustive exploration of parameter space in this work. The behavior we show here is generally representative of the full parameter space, and most parameters result in relatively trivial changes in the behavior (i.e., scaling either the luminosity or time axes, without changing the variability structure).  We make the stochastic light curve model publicly available with this manuscript\footnote{link will go here.}, and encourage experimentation and comparison to observations. 

The typical behavior of these stochastic light curves is shown in Figure \ref{fig:long}. For which we take a correlation time $t_{\rm corr} = 5$ days, a peak of the mean disk temperature $k \hat \mu_T = 120$ eV, and fix $L_0$ and $t_{\rm evol}$ so that the peak of the mean X-ray luminosity $L_{X, {\rm peak}} \sim 10^{43}$ erg/s and decay rate ($\sim 2$ orders of magnitude in two years) are similar to a typical TDE. We fix $\eta=2$. We then vary the fractional temperature variability $\sigma_T/\mu_T$, and compute various properties. Each system is assumed to be ``observed'' every day for two years. In the upper panels we display some random realisations of long time stochastic X-ray light curve evolution for two different values of the fractional temperature variability $\sigma_T/\mu_T = 0.05$ (left) and $\sigma_T/\mu_T = 0.1$ (right). It is clear to see from this Figure that the observed X-ray luminosity variability is a sensitive function of assumed temperature variability. Note that $\sim$ order of magnitude short-timescale X-ray dimming events are seen in the right hand panel, as has also been seen in tidal disruption events.

In the lower panel we display some statistical properties of these long-timescale stochastic systems, computed using a large sample of realisations. These properties were chosen as they are most readily compared to observations. On the left we show the distribution of the local luminosity ratio (i.e., $R=l(t)/\overline L(t)$, where $l(t)$ is a stochastic light curve realisation and $\overline L(t)$ is the mean evolution defined above) as a function of the assumed fractional temperature variability. On the right we show a large sample of the so-called flux-RMS relationship predicted in this theory, for three different values of the  fractional temperature variability. Note that the fractional variability in the luminosity can reach $\sigma_L/\mu_L\sim 1$ for $\sigma_T/\mu_T \sim 0.1$, an exponential enhancement. 

This RMS-flux relationship is of intrinsic interest, and is a commonly observationally derived property of accreting systems \citep[see e.g.,][among many others]{Uttley05}. We see that the theory developed here predicts that TDEs will themselves have {\it an} RMS-flux relationship (i.e., $\sigma_L$ and $\mu_L$ are indeed  correlated), but that this relationship will be slightly sub-linear (note that a linear relationship is a property of a log-normal distribution, and this is therefore potentially interesting). However, in the much smaller observational sample sizes available to real observations of TDE systems (i.e., a at best a $\sim$ hundred points in a lightcurve) it seems unlikely that any deviation from linear will be detectable at high significance given the intrinsic  uncertainties inherent to X-ray observations, but we believe this is worth testing observationally. 

The lower left panel highlights the natural degeneracy between $\sigma_T/\mu_T$ and $t_{\rm corr}$ discussed earlier (see also the lower right panel of Figure \ref{fig:light_curves}). The key point is that if one has access to a long-term X-ray light curve of a TDE, then the effects of $t_{\rm corr}$ are washed out as one naturally probes observational windows $\Delta t \gg t_{\rm corr}$. Therefore these long time light curves are better probes of of $\sigma_T/\mu_T$ which, once constrained, would allow short timescale observations to probe the turbulent correlation time $t_{\rm corr}$. 

\section{Comparison to observations}\label{obs}
\begin{figure}
    \centering
    \includegraphics[width=0.48\linewidth]{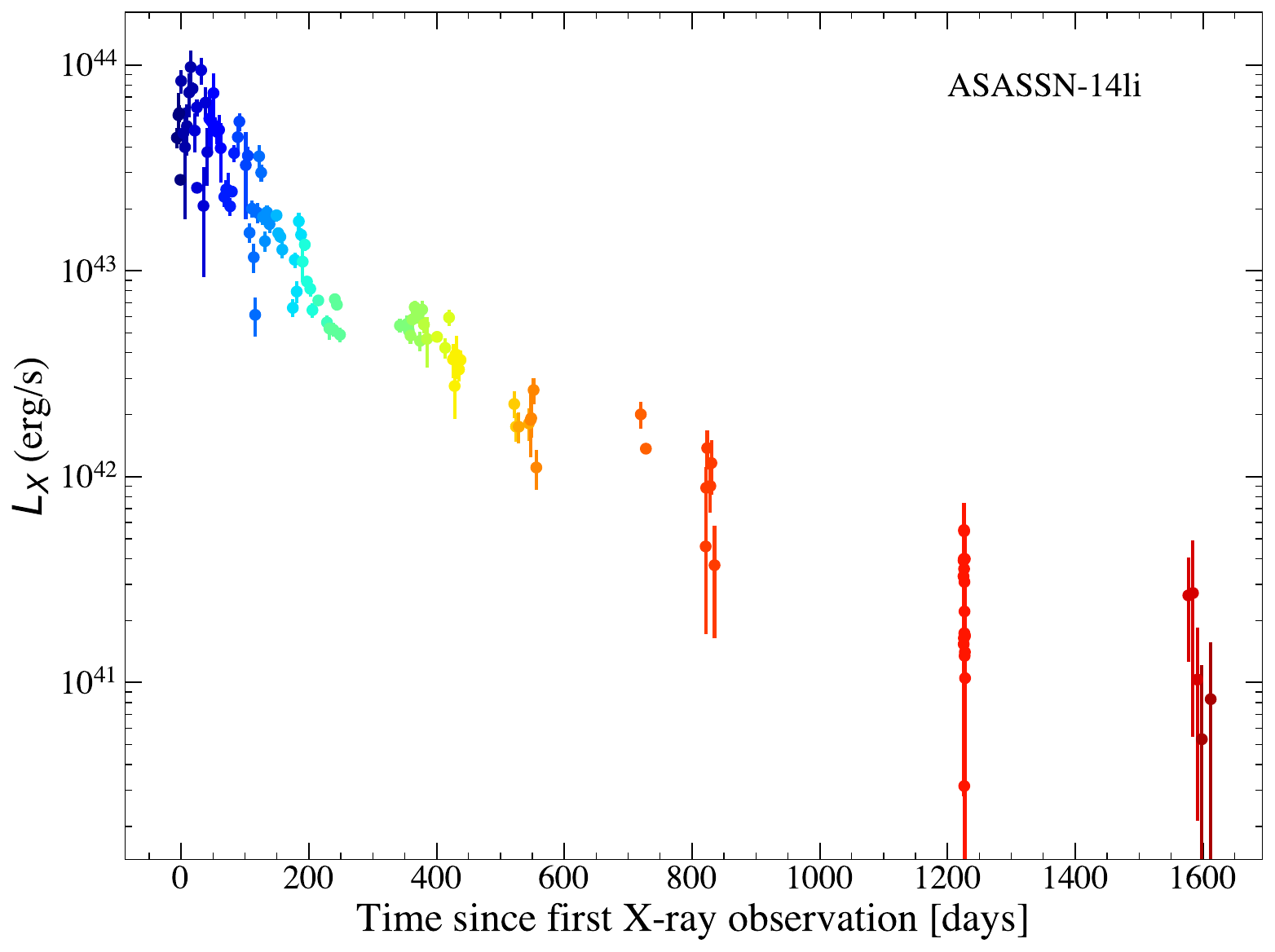}
    \includegraphics[width=0.48\linewidth]{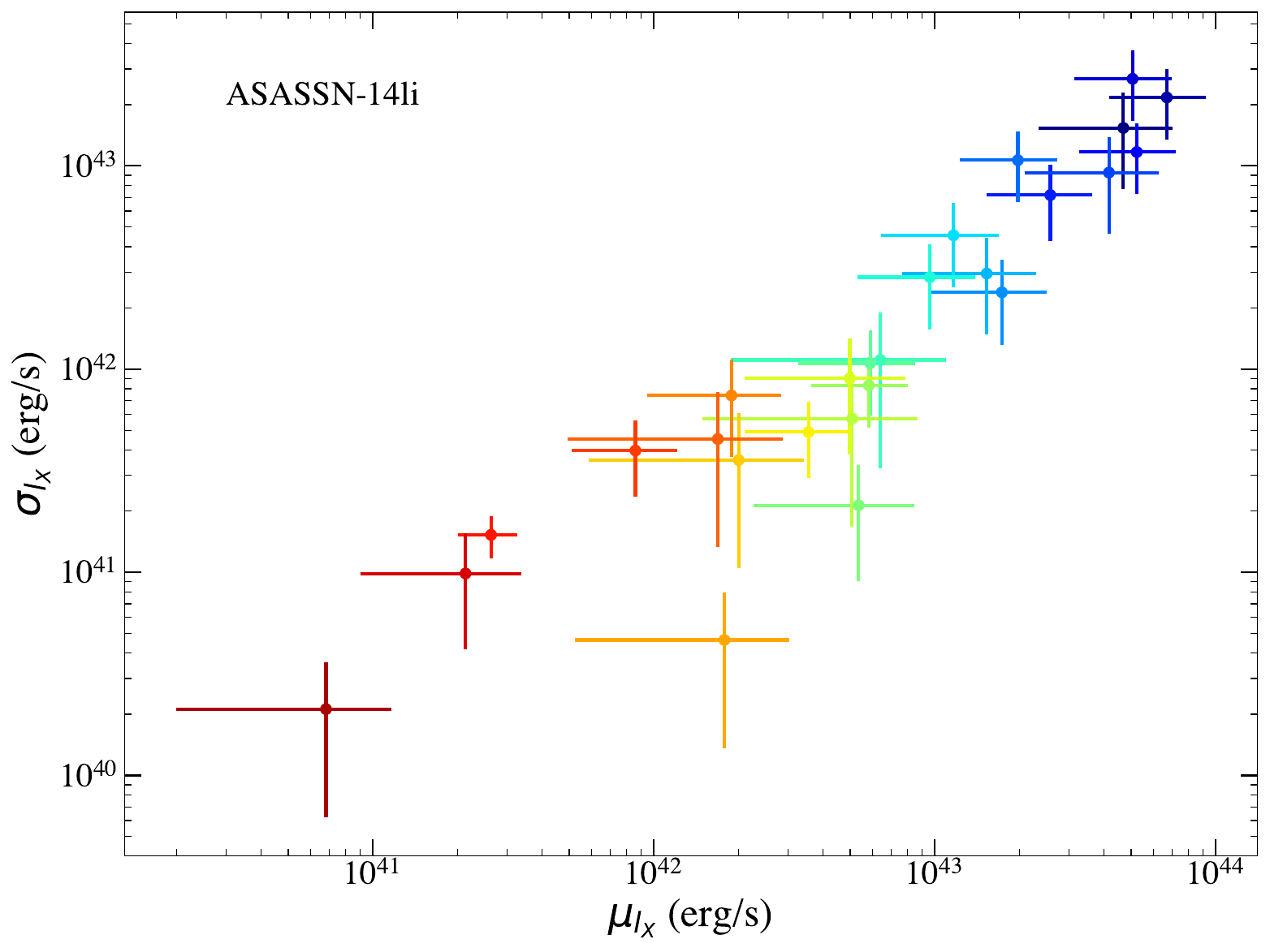}
    \includegraphics[width=0.48\linewidth]{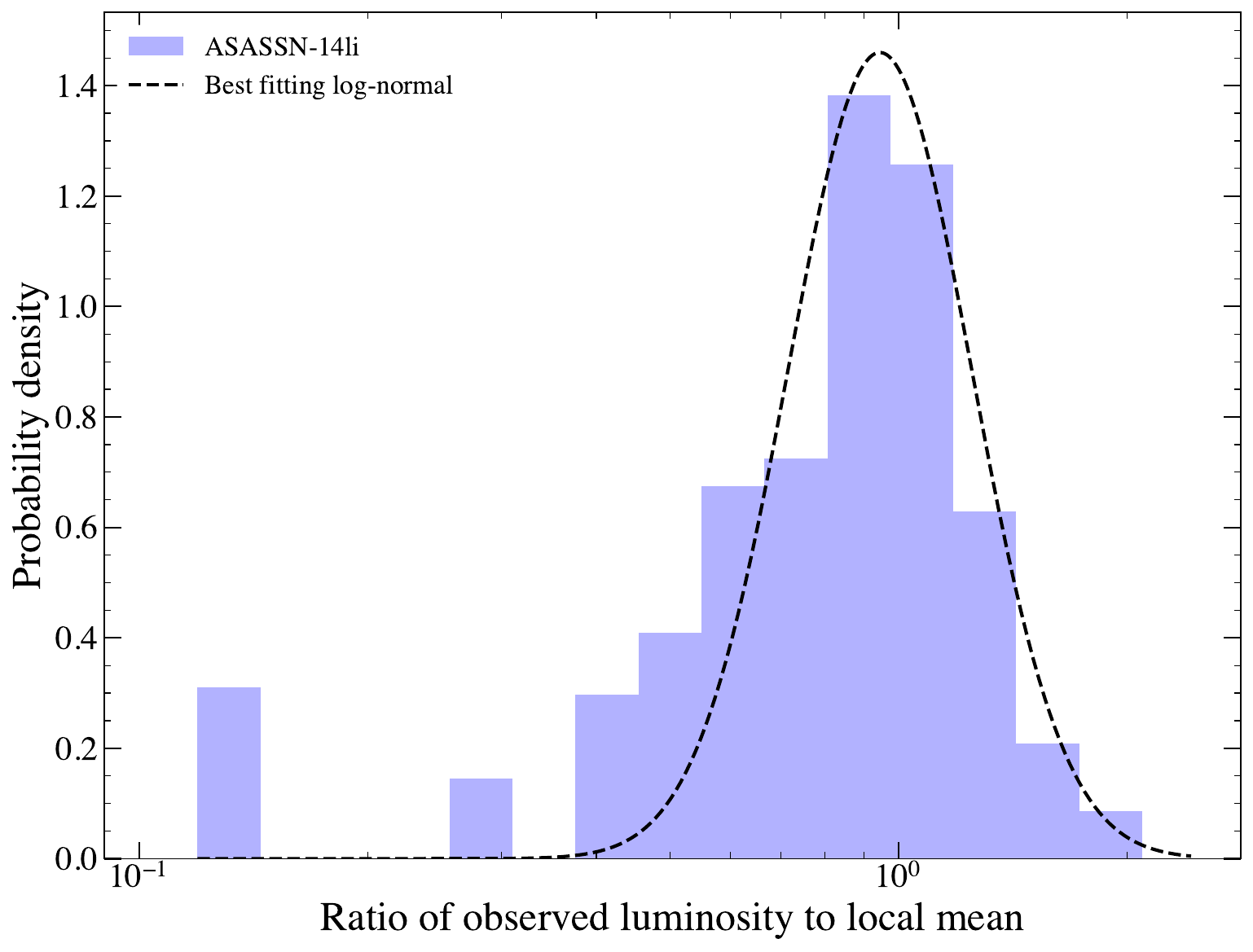}
    \caption{The observed X-ray light curve of the TDE ASASSN-14li ({\bf upper left}), with X-ray data binned into time periods spanning $\Delta t=21$ days (denoted by the color of the points on each panel). There is clear variability about a smoothly evolving mean, and the standard deviation of the intra-bin X-ray emission is plotted against the mean of the bin in the {\bf upper right} panel (this is the so-called RMS-flux relationship). Clearly, ASASSN-14li follows a near-linear RMS-flux relationship, with the standard deviation of the luminosity falling off with the mean.  This does not mean that the X-ray luminosity of ASASSN-14li is log-normally distributed however, as the variation in the evolving luminosity about its mean is poorly described by a log-normal distribution ({\bf lower panel}).   }
    \label{fig:14li_data}
\end{figure}

In this final section we contrast the predictions of the theoretical framework developed in this paper to X-ray observations of two tidal disruption events with notably distinct variability structures, namely ASASSN-14li \citep{Miller15} and AT2022lri \citep{Yuhan24}. These two sources also happen to the two brightest (non-jetted) TDEs observed in the last decade at X-ray energies (a combination of their intrinsic luminosity and relative proximity). 

In the first two sub-sections we consider only the {\it long timescale} evolution and variability of these systems (i.e., we do not consider the variability across a given observation). Computing and analyzing this intra-observation variability is more technically involved, and we perform a qualitative analysis in section 5.3, with further analysis postponed for a detailed follow up study. 

We have specifically chosen these two sources as they have X-ray data which is (i) well sampled, (ii) covers a long baseline and (iii) show clearly different (and interesting) variability structure. The X-ray spectra  of both of these sources are dominated by thermal emission \citep{Miller15, Yuhan24}, with (energetically) small outflow features, making them ideal targets for a pure disk theory analysis (i.e., neither has strong emission from a corona which would complicate the analysis\footnote{It should be noted that high sensitivity X-ray observations of both sources do show some features which are not purely-thermal \citep{Kara18, Yuhan24}, these features however make up a small fraction of the total X-ray luminosity and therefore should be sub-dominant in terms of the variability.}). 

\subsection{ASASSN-14li}
ASASSN-14li is a well studied \citep[e.g.][]{Miller15, Holoien16b, MumBalb20a, Wen20, GuoloMum24} X-ray and optically bright TDE, with over 1600 days of X-ray data, and over 3000 days of UV data, both well described by thermal disk emission \citep{Guolo24, GuoloMum24}. This TDE therefore represents an ideal first test for the stochastic light curve theory developed here. 

\begin{figure}
    \centering
    \includegraphics[width=0.48\linewidth]{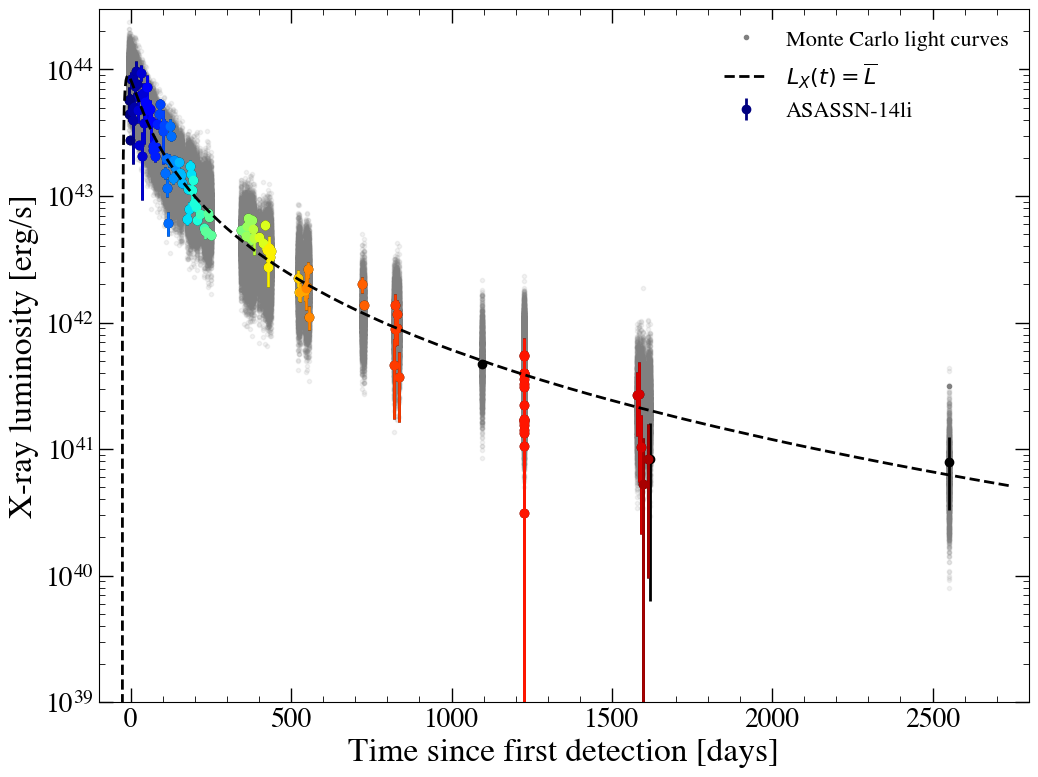}
    \includegraphics[width=0.48\linewidth]{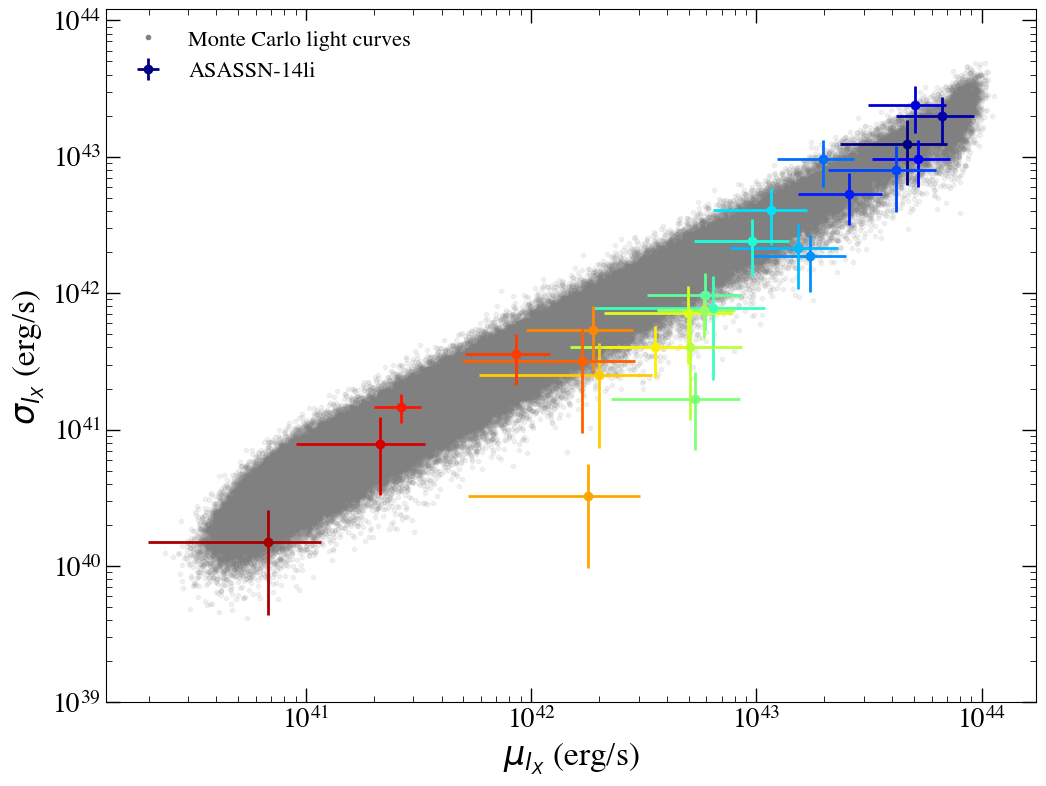}
    \includegraphics[width=0.48\linewidth]{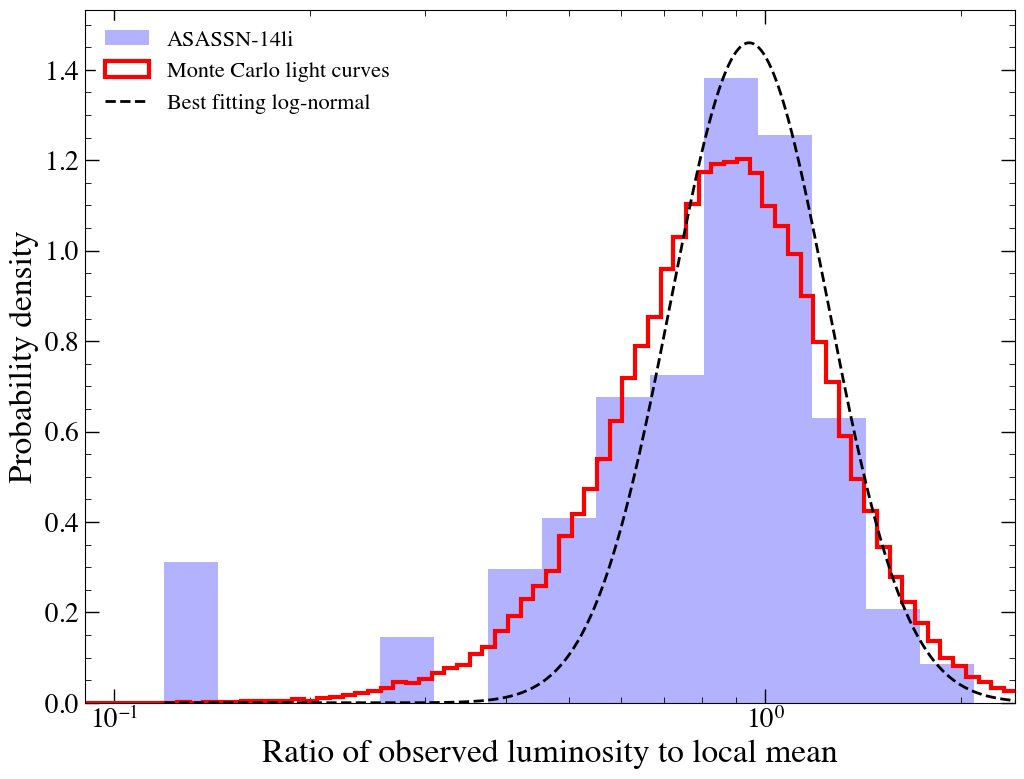}
    \caption{A comparison of the X-ray evolution of ASASSN-14li to the theoretical model developed here. The data are reproduced from Figure \ref{fig:14li_data}, and we now add the predictions of a Monte Carlo sampling of $N = 2000$ stochastic light curve systems with fractional temperature variability $\sigma_T/\mu_T = 0.06$ (see text for a description of the other parameters). In the {\bf upper left} plot we show by gray points each of the sampled luminosities, and the mean evolution of the system with $\sigma_T = 0$ (black dashed curve). The Monte Carlo sample of the predicted RMS-flux relationship is shown in the {\bf upper right} panel, which clearly reproduces the trend seen in the ASASSN-14li data.  The predicted distribution of the disk luminosity about its mean is shown in the {\bf lower} panel, where the theoretical model developed here clearly provides a vastly improved description of the observations when compared to a log-normal distribution.   }
    \label{fig:14li_results}
\end{figure}

In Figure \ref{fig:14li_data} we show the evolving X-ray light curve of ASASSN-14li (upper left panel), with the data grouped (by colour) into blocks spanning $\Delta t = 21$ days (only blocks with $N>3$ observations in this time period are shown). The data is taken from \cite{Guolo24}, and is predominantly made up of {\it Swift} XRT 0.3-10 keV observations, with a small number of {\it XMM-Newton} 0.3-10 keV observations included.  This bin size was chosen as a compromise between having enough bins to see the evolution of $\sigma_L$ and $\mu_L$ with time, without having too many bins with a small number of points.  It is clear that the mean evolution of ASASSN-14li is smooth, and indeed this bulk evolution is well described by an evolving relativistic disk \citep{MumBalb20a, GuoloMum24}, as we shall highlight again shortly. 

There is also, however, clear variability about this mean evolution. In the upper right panel we show the intra-bin mean and standard deviation of the ASASSN-14li X-ray luminosity (i.e., the flux-RMS relationship of this source). Again, the time of each bin is denoted by the colour of the data point. We use a fractional uncertainty on both $\mu_L$ and $\sigma_L$ set by $1/\sqrt{N_{{\rm bin}, i}}$, where $N_{{\rm bin}, i}$ denotes the number of points in temporal bin $i$. It is clear that ASASSN-14li clearly displays {\it a} flux-RMS relationship, and indeed the data shown in the upper right panel are adequately described by a linear relationship (i.e., $\log \sigma_L = \log\mu_L + B$). 

While a linear relationship in the $\log \sigma_L - \log\mu_L$ plane could be interpreted as evidence for a log-normally distributed luminosity, this is emphatically not the case for the X-ray light curve of ASASSN-14li. This can clearly be seen in the lower panel of Figure \ref{fig:14li_data}, where we plot the distribution of the local luminosity ratio of the light curve, defined by 
\begin{equation}
R = {\{L_j\}\over \left\langle \{L_j\}\right\rangle} , \,\, {\rm for} \,\, \{L_j\}=L_{\rm obs}(t_j)\,\, {\rm for} \,\,t_j \in i^{\rm th} \,\,{\rm time}\,\, {\rm bin},
\end{equation}
and $\left\langle \cdot \right\rangle$ denotes a simple average within the same bin. 

This quantity is equal to the variability in the light curve {\it about} the evolving mean of the light curve. We show the best-fitting log-normal distribution to this $p_R$ distribution by the black dashed curve in this lower panel, demonstrating that this is a poor description of the system. 

The reason for this is that a log-normal distribution does not produce enough {\it dimming} events, which are observed in the ASASSN-14li light curve. The over-abundance of dimming events is precisely a prediction of the theory developed in this paper (see Figure \ref{fig:evs} and section \ref{evs}).

Indeed, the theory developed here is able to reproduce the variability structure of ASASSN-14li, as we demonstrate in Figure \ref{fig:14li_results}. In Figure \ref{fig:14li_results} we reproduce the data from Figure \ref{fig:14li_data}, but now include the predictions of the stochastic light curve theory. 

We first fix the gross disk properties by reproducing the mean evolution of the X-ray light curve. The peak mean temperature $\hat \mu_T$ was set to that inferred from fits to the X-ray spectra \citep[e.g.,][]{Brown17, Mummery_Wevers_23, Guolo24}, and then the luminosity amplitude $L_0$ was set to reproduce the observed peak luminosity (this corresponds to a black hole mass $M_\bullet \approx {\rm few} \times 10^6 M_\odot$, with significant uncertainty related to the spin and inclination of the system, but broadly in line with other measurements). With the peak luminosity reproduced, we modified the evolutionary timescale $t_{\rm evol}$ with $\sigma_T=0$ to reproduce the gross evolution of the light curve, before letting $\sigma_T/\mu_T$ vary from zero. We fixed $\eta = 2$ for this analysis, assuming a face-on disk. We find minimal sensitivity to the correlation timescale for this long timescale evolution, provided that it was short compared to the gross evolution (i.e., any choice $t_{\rm corr} \lesssim 15$ days reproduced the data, the plot explicitly takes $t_{\rm corr} = 1$ day\footnote{This is simply a result of the fact that the observational cadence sets the timescale over which one is sensitive to correlations, and the theory is insensitive to correlation timescales shorter than $\sim$ the cadence.}). In reality, there are variety of choices of $L_0, \hat \mu_T, \eta$ and $t_{\rm evol}$ which broadly reproduce the mean evolution, and do not change the quantitative results presented here.  This mean evolution is shown by the black dashed curve in the upper left panel of Figure \ref{fig:14li_results}. 

We then varied $\sigma_T/\mu_T$ (but kept the ratio fixed as a function of time), finding that $\sigma_T/\mu_T = 0.06$ provided a good description of the variability features. We have not performed a detailed statistical fit here, merely judging by eye what value of the ratio reproduces the broad features displayed in Figure \ref{fig:14li_data}. 

In the upper left plot of Figure \ref{fig:14li_results} we show the observed X-ray light curve of ASASSN-14li (coloured points), the mean evolution (black dashed curve) and $N=2000$ realisations of stochastic light curves evaluated at the same times as the observations taken of ASASSN-14li (grey dots). In other words we Monte Carlo simulate a large number of realisations of what we believe to be the underlying stochastic system. We see that the variability around the mean evolution can clearly be accounted for with a simple log-normally distributed disk temperature,  with moderate fractional variability, and the observational effects of observing within the Wien tail.  

A more detailed comparison is shown in the upper right, and lower, panels of Figure \ref{fig:14li_results}. In the upper right panel we show a Monte Carlo simulation of $N=2000$ realisations of the flux-RMS relationship for ASASSN-14li. Clearly, the observations are consistent with the simplified stochastic model developed here. We note that the outliers (on the flux-RMS plot) typically result from light curve bins with a small number of points. In the lower plot we show (in red) the distribution of the stochastic light curves around their mean evolution (here explicitly this is $\overline L(t)$, the black dashed curve in the upper left panel). Again, this is a Monte Carlo simulation of $N=2000$ realisations of the light curves. This distribution provides a significantly improved description of the observed ASASSN-14li distribution, and in particular reproduces the asymmetry of the distribution, with the model showing a similar propensity for dimming events as the data. As we have argued throughout this paper this asymmetry is a result of observing ASASSN-14li in the Wien tail, a fact we demonstrate explicitly with numerical experimentation in Appendix \ref{app:C}. 

While the improvement over a log-normal description of the variability is clear by eye, we can indeed quantify this improvement more robustly. The likelihood ${\cal L}$ quantifies the probability of seeing the a data set $\{R_i\}$ given an assumed probability density function $p_R(R|{\rm model)}$, which is of course model dependent (in our case this is in fact the model itself). This likelihood function can be used to distinguish between different models of the underlying variability. Formally, the likelihood is 
\begin{equation}
    {\cal L}({\rm model}) \equiv p({\rm data}|{\rm model}) = \prod\limits_{i=1}^{n} \,p_R(R_i|{\rm model}), 
\end{equation}
or equivalently, and more usefully, the log-likelihood is equal to 
\begin{equation}
    \ln {\cal L}({\rm model}) = \sum\limits_{i=1}^{n} \, \ln p_R(R_i|{\rm model}) .
\end{equation}
While the actual values of $\ln {\cal L}$ for any given model are not of interest, the difference between the log-likelihoods for two different models is, as it quantifies an improvement in the description of the variability. The so-called Bayesian information criteria is defined (for two models with the same number of free parameters, which is the case here) as
\begin{equation}
    \Delta {\rm BIC} = 2\ln {\cal L}({\rm model}') - 2\ln {\cal L}({\rm model}) = 2\sum\limits_{i=1}^{n} \, \ln p_R(R_i|{\rm model'}) - \ln p_R(R_i|{\rm model}) ,
\end{equation}
and the following ``rule of thumb'' states that $\Delta {\rm BIC} > 10$ shows very strong evidence in favor of ${\rm model}'$ over ${\rm model}$ \citep[see e.g.,][for a more rigorous relationship between $\Delta {\rm BIC}$ and the evidence for a model]{Trotta08}. 

For ASASSN-14li we find that the best fitting log-normal has $\ln {\cal L}_{\rm LN} = -71.5$, while the model derived in this paper has $\ln {\cal L}_{\rm MB}=-42.4$, meaning that $\Delta {\rm BIC} = 58.4$, far exceeding the rule of thumb value. We believe that ASASSN-14li provides clear support for the model developed here. 

\subsection{AT2022lri}
\begin{figure}
    \centering
    \includegraphics[width=0.48\linewidth]{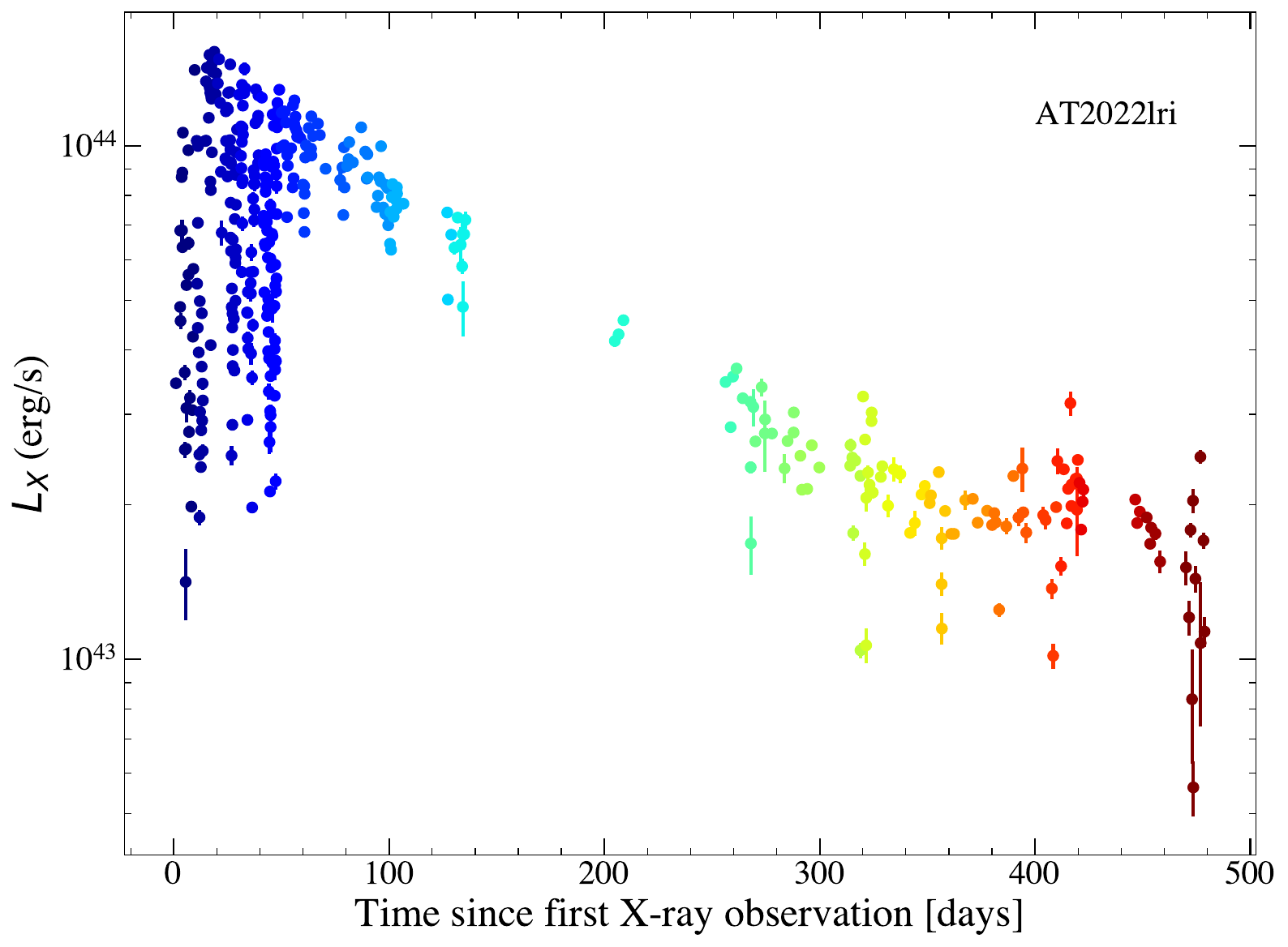}
    \includegraphics[width=0.48\linewidth]{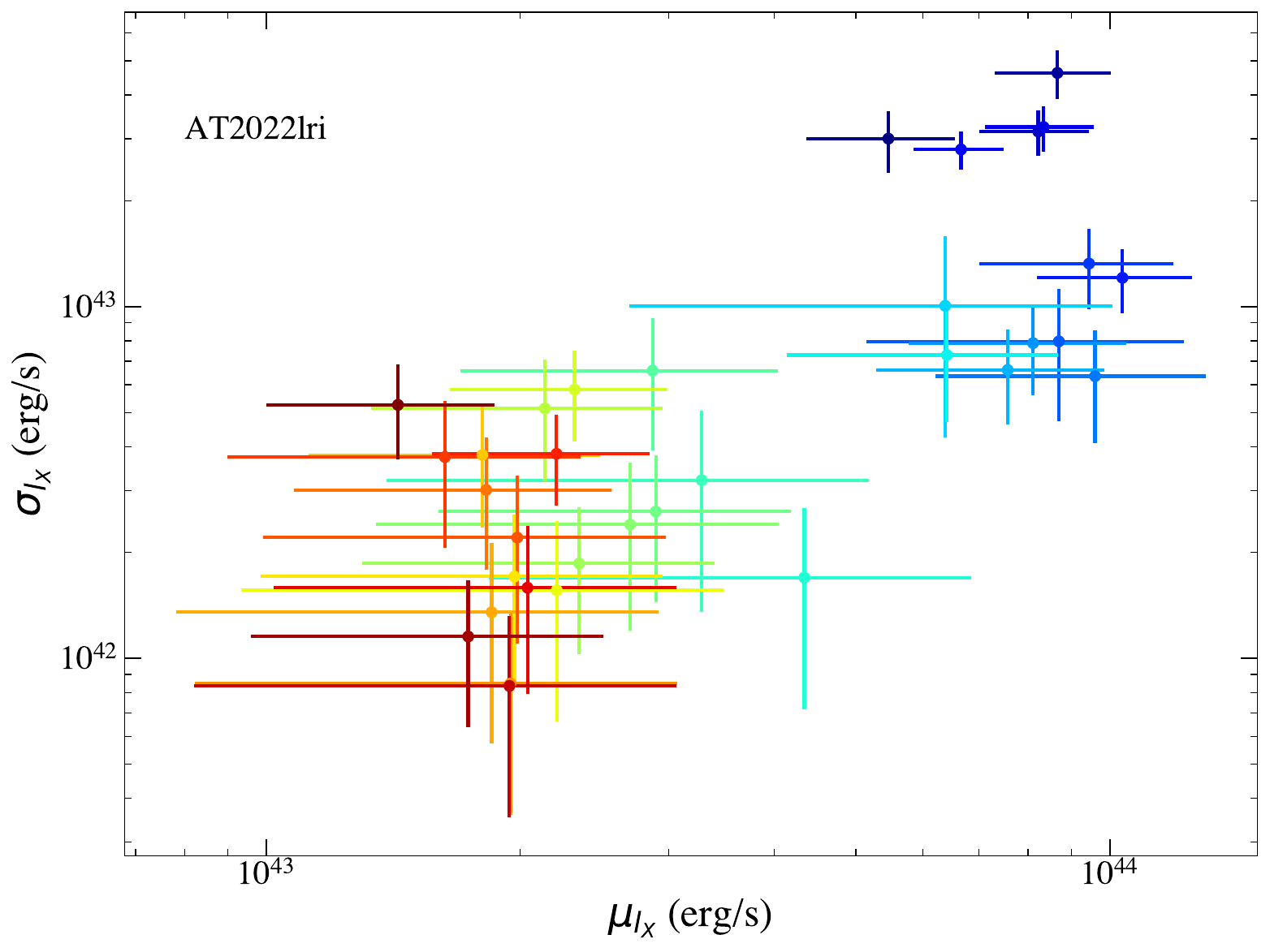}
    \includegraphics[width=0.48\linewidth]{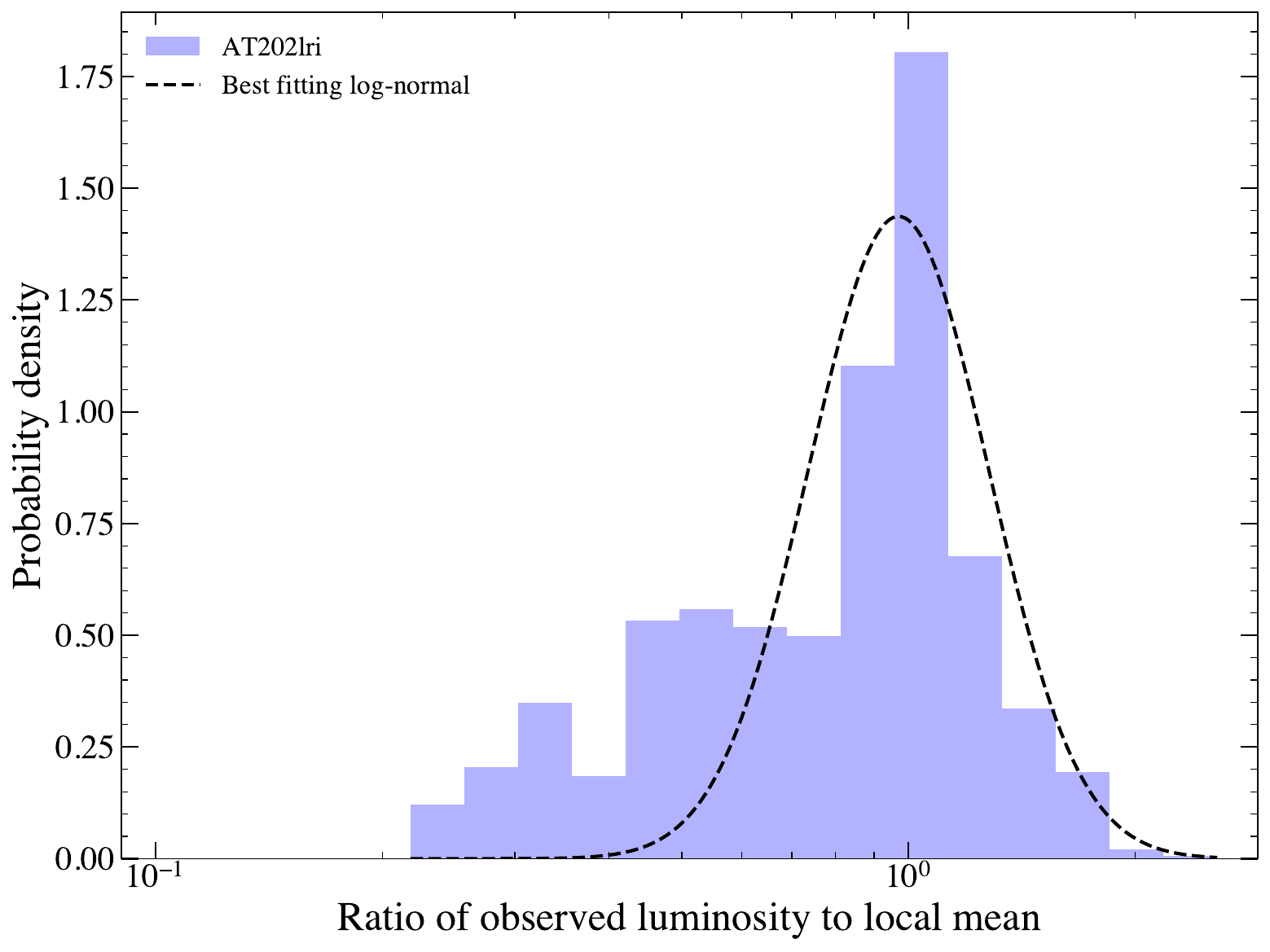}
    \caption{A very similar figure to Figure \ref{fig:14li_data}, except now for the TDE AT2022lri, which shows more interesting variability structure. In the {\bf upper left} panel we show the X-ray light curve of AT2022lri, with X-ray data binned into time periods spanning $\Delta t=10$ days (denoted by the color of the points on each panel). There is clear variability about a smoothly evolving mean, and most interestingly there is a qualitative change in the variability structure at $t \sim 60$ days, with a sudden drop in fractional X-ray variability. The standard deviation of the intra-bin X-ray emission is plotted against the mean of the bin in the {\bf upper right} panel (this is the so-called RMS-flux relationship). AT2022lri follows a near-linear RMS-flux relationship for times $t \gtrsim 60$ days, but has a higher variance in its X-ray luminosity (at fixed mean) for $t \lesssim 60$ days.  The later near-linear RMS-flux relationship does not mean that the X-ray luminosity of AT2022lri is log-normally distributed however, as the variation in the evolving luminosity about its mean is poorly described by a log-normal distribution ({\bf lower panel}).  }
    \label{fig:22lri_data}
\end{figure}

While ASASSN-14li shows relatively simple stochastic variability, the more recently discovered TDE AT2022lri \citep{Yuhan24} shows more complex, and therefore more interesting, variability. In Figure \ref{fig:22lri_data} we show the same simple  plots as in the earlier ASASSN-14li analysis, simply now with the X-ray light curve of AT2022lri \citep[data taken from][and was taken using the {\it NICER} instrument, again with an energy range of 0.3-10 keV]{Yuhan24}. The X-ray light curve of AT2022lri was sampled at a higher observational cadence, and so we bin this light curve on a shorter timescale $\Delta t_{\rm bin} = 10$ days.

Something which is clear from a visual inspection of Figure \ref{fig:22lri_data} is that the evolution of the variability in AT2022lri is non-trivial, with a clear switch to a lower (but clearly non-zero) variability state at a time roughly $\Delta t \sim 60$ days after first detection.  This can be seen both visually in the light curve itself, but also in the $\sigma_L-\mu_L$ plot (upper right), which displays a multi-valued behavior, with $\mu_L\sim 10^{44}$ erg/s corresponding to two different variability states $\sigma_L\sim 10^{43}$ erg/s and $\sigma_L \sim 5\times 10^{43}$ erg/s. Again, much as was the case for ASASSN-14li, AT2022lri is poorly described by a log-normal distribution (lower panel), as a log normal is unable to explain the large number of dimming events observed in the light curve of AT2022lri. 

\begin{figure}
    \centering
    \includegraphics[width=0.48\linewidth]{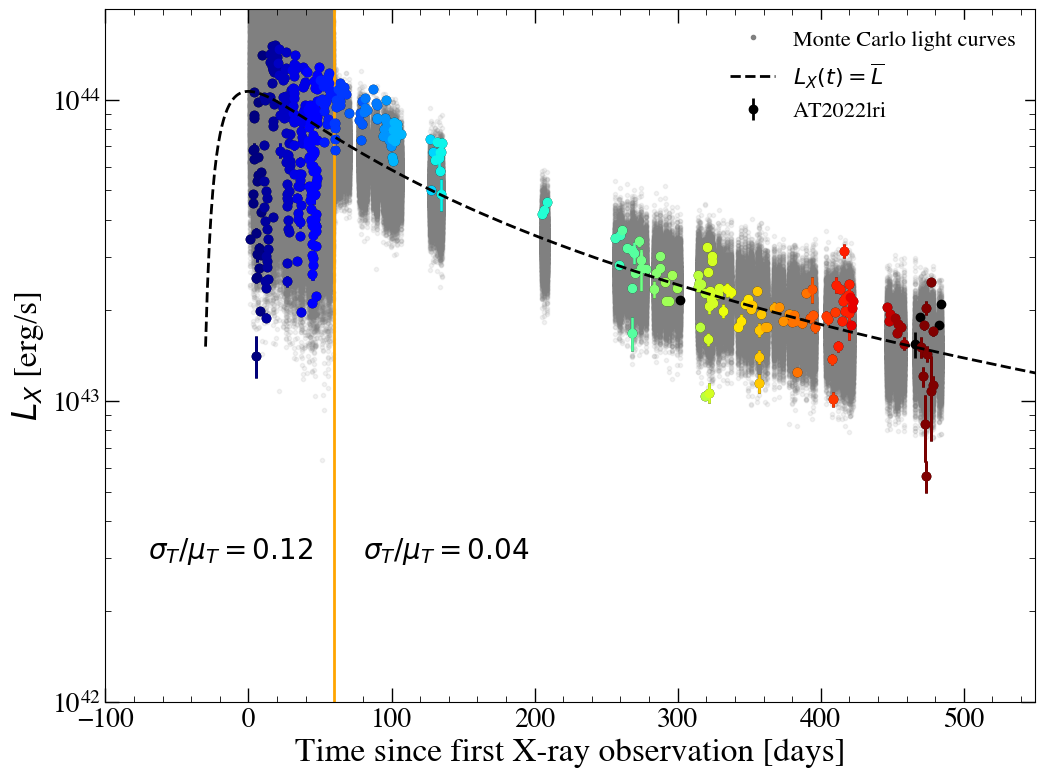}
    \includegraphics[width=0.48\linewidth]{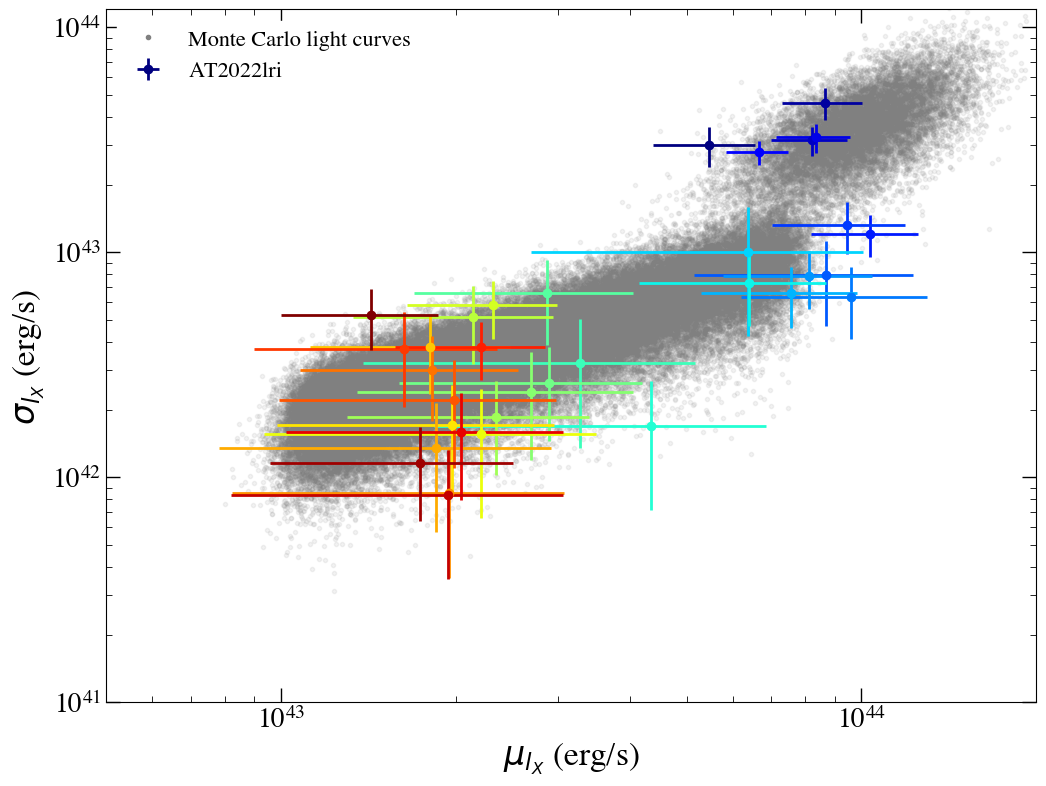}
    \includegraphics[width=0.48\linewidth]{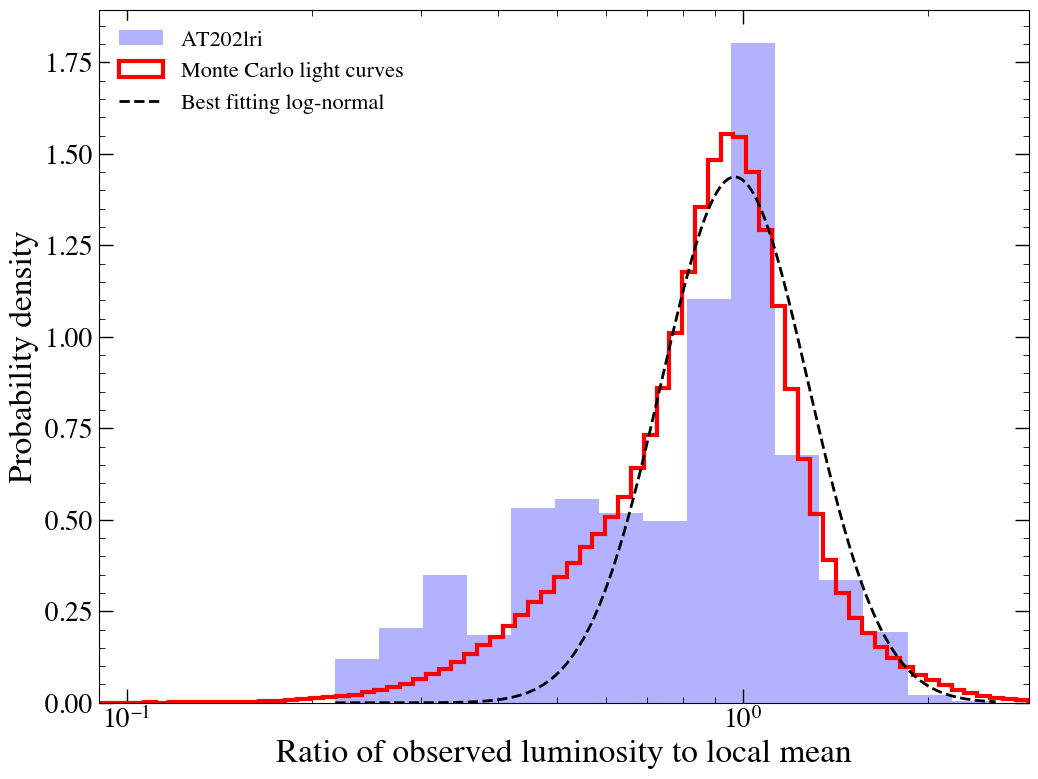}
    \caption{A comparison of the X-ray evolution of AT2022lri to the theoretical model developed here. The data are reproduced from Figure \ref{fig:22lri_data}, and we now add the predictions of a Monte Carlo sampling of $N = 2000$ stochastic light curve systems with fractional temperature variability $\sigma_T/\mu_T$ that undergoes a sharp transition at $t \sim 60$ days (see text for a description of this and the other parameters). In the {\bf upper left} plot we show by gray points each of the sampled luminosities, and the mean evolution of the system with $\sigma_T = 0$ (black dashed curve). The orange vertical line shows the time at which the model transitions between variability states.  The Monte Carlo sample of the predicted RMS-flux relationship is shown in the {\bf upper right} panel, which clearly reproduces the trend seen in the AT2022lri data, including the two-branch structure.  The predicted distribution of the disk luminosity about its mean is shown in the {\bf lower} panel, where the theoretical model developed here clearly provides a vastly improved description of the observations when compared to a log-normal distribution.  }
    \label{fig:22lri_results}
\end{figure}

To reproduce the mean evolution of AT2022lri we followed an identical procedure to that discussed above for ASASSN-14li. However, the change in variability state of AT2022lri clearly means that a simple $\sigma_T/\mu_T = {\rm constant} \,\, \forall t$ model will clearly not reproduce the data. To model this change in observed variability we perform a particularly simple modeling choice, by taking 
\begin{equation}
    \sigma_T/\mu_T = \begin{cases}
        F_1, \quad t < 60\, {\rm days}, \\
        \\
        F_2, \quad t \geq 60 \, {\rm days},
    \end{cases}
\end{equation}
after which we then vary $F_1$ and $F_2$. We first perform this as a pure modeling exercise, and postpone a discussion of the possible physics of this variability transition to a later point. 

The results of this analysis are shown in Figure \ref{fig:22lri_results}, for $F_1 = 0.12$ and $F_2 = 0.04$. Again all plots are made for a Monte Carlo sampling of $N = 2000$ realisations. The early time behavior of the light curve itself is hard to discern, given the high variability state the system is in at that moment. The more relevant comparisons are to the $\sigma_L-\mu_L$ plot in the upper right panel, and to the distribution of the luminosity about its mean in the lower panel. 

As is clear form all three panels, the lower variability (more ASASSN-14li like, $\Delta t \geq 60$ days) state of AT2022lri is well described by the theory developed here. This is most clearly seen in the lower branch of the $\sigma_L-\mu_L$ plot (upper right panel), which is well reproduced. Adding in an early time, higher variability, state reproduces the upper branch of the $\sigma_L-\mu_L$ plot, and the large amplitude dimming events seen in the light curve itself. We note at this point that the early-phase model predicts a moderate probability of some higher luminosity flares at early times not seen in the data. This may be chance in a single realisation of a stochastic process (see for example the distribution in the lower panel), or it could be a result of missing physics in the model. We note that the Eddington limit is not explicitly incorporated into the theoretical framework considered here, and it is known that this bolometric luminosity scale limits the X-ray luminosity to $L_X \lesssim 10^{44}$ erg/s \citep{Mum21}, a theoretical prediction with strong observational support \citep{Guolo24, Grotova25}. It is possible that X-ray flares are further curtailed by global Eddington-limit constraints, in addition to the asymmetry introduced by the Wien-tail which we model here. 

The joint distribution of the two phases about their mean evolution also provides a good description (and significantly better than the log-normal) of the data (lower panel, red curve). Again, by using the Bayesian information criterion we can distinguish between these two models. The criterion is modified as the two models no longer have the same number of free parameters, as we have added a free parameter to the disk theory (the change in fractional temperature variability). In the BIC formalism this leads to a penalization of $\ln n$ times the difference in the number of free parameters, or in our case 
\begin{equation}
    \Delta {\rm BIC} = 2\ln {\cal L}({\rm model}') - 2\ln {\cal L}({\rm model}) - \ln(n) = - \ln(n) + 2\sum\limits_{i=1}^{n} \, \ln p_R(R_i|{\rm model'}) - \ln p_R(R_i|{\rm model}) ,
\end{equation}
as ${\rm model}'$ has one more free parameter than ${\rm model}$. We find that the best fitting log-normal has $\ln {\cal L}_{\rm LN} = -278.5$, while the model developed here has $\ln {\cal L}_{\rm MB} = -144.0$, and therefore, despite $n = 424$ and $\ln n \approx 6$, we have $\Delta {\rm BIC} = 263$, which again counts as extremely strong evidence in support of the framework developed here. 

The question, naturally, is what physical process may cause such a change in variability state. We speculate about possible mechanisms in the final section of this manuscript. 

\subsection{Short (intra-observation) timescale variability}

The results of the previous two subsections examined the variability of two TDEs on long (hundreds of days) timescales. The analysis of short (intra-observation) variability is more  complicated (on a technical level), as one must take into account the energy dependence of the response of the instrument, which can impact the number of counts ``seen'' at different energies. 

We do not therefore attempt a quantitative comparison of the theory we have developed here with intra-observation variability. Instead, we perform a qualitative examination of the main prediction of this theory on short timescales, namely that the fractional variability of the observed emission will increase with the energy at which the source is observed across a band. We stress that future, detailed, analysis and comparisons are required to move beyond the simple qualitative approach undertaken here, and that this would be an ideal future test of classical disk theory. 

\begin{figure}
    \centering
    \includegraphics[width=0.95\linewidth]{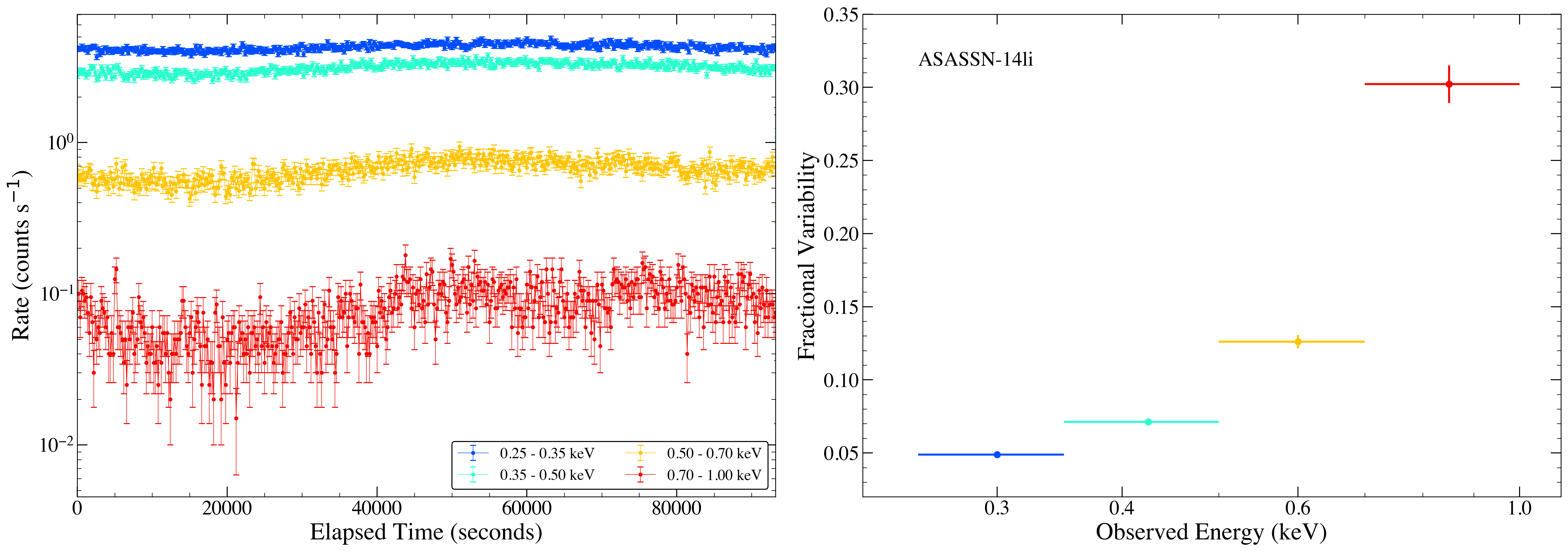}
    \includegraphics[width=0.95\linewidth]{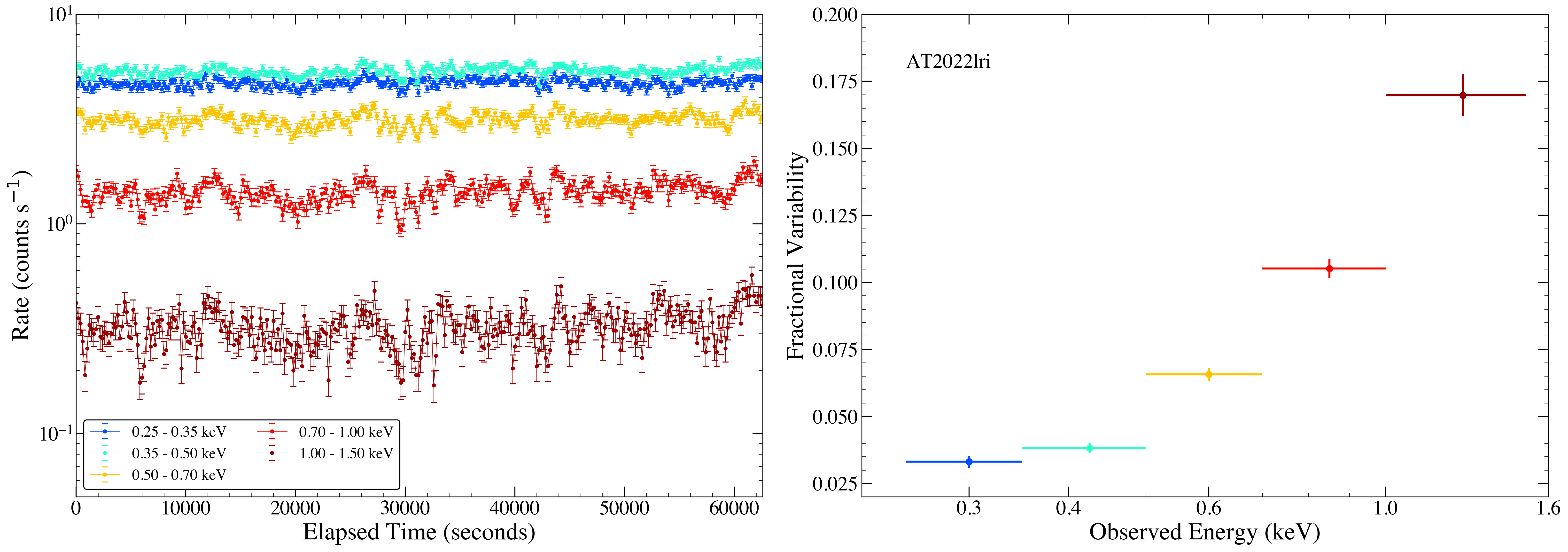}
    \caption{The short timescale (intra-observation) energy dependent variability of ASASSN-14li (upper) and AT2022lri (lower). The light curves in different energy bands are shown on the left, with the fractional variability displayed on the right. The light curves show a visible increase in variability with increased observing energy, which is confirmed by explicit computation. This result is strongly supportive of the theory developed here. }
    \label{fig:short_time}
\end{figure}

In Figure \ref{fig:short_time} we show a single time resolved XMM observation of ASASSN-14li (top panel), and a single time resolved XMM observation of AT2022lri\footnote{I am indebted to Muryel Guolo who reduced and then provided the data for both sources used in this Figure. }. On the left we display the observed count rate (the number of photons detected per second) split across different energy bands of each instrument as a function of time. The fractional variability in each band is then plotted against the energy range of the band on the right hand panel. It is clear from an inspection of the light curve itself that the uncertainty in each measured count rate increases with energy (a natural result of higher energy emission being fainter, and therefore its observed value more uncertain, in the Wien tail). To ensure that an inference of an increased fractional variability is not a result of this increased measurement uncertainty we require an estimate of the uncertainty in $F_C = \sigma_C / \mu_C$ (where $C$ denotes count rate). We do this by following the classic text of \cite{Vaughan03}, and compute 
\begin{equation}
    \delta F_C = \sqrt{{1\over 2N}\left({\hat \sigma^2 \over \hat \mu^2 F_c}\right)^2 + {\hat \sigma^2 \over N (\hat \mu)^2}},
\end{equation}
which is asymptotically equal to the uncertainty in $F_C$ in the large data set size $(N)$ limit. In the above expression $\hat \sigma^2 = N^{-1}\sum_i\sigma_i^2$, $\hat \mu = N^{-1}\sum_i C_i$ and  $\hat \mu^2 = N^{-1}\sum_i C_i^2$ where $\sigma_i$ is the uncertainty in any one measurement of the counts $C_i$. 

It is clear that the data from  both sources is in strong qualitative agreement with the predictions of the theory developed here, which can be seen in particular by reference to Figure \ref{fig:light_curves}. Both a visual inspection and comparison of the predicted and observed light curves, but also the clear increase in fractional variability with energy offer (strong) support for the theory developed here. This increase in fractional variability is intrinsic, and is a not a result of increased measurement uncertainty in the fainter regions of the spectrum. 

We stress that the above results can only be taken to be supportive of the model developed here, and future analysis is certainly warranted to quantitatively test these models over the shortest timescales. This analysis will need to take into account the energy dependent instrument response but, we suspect, appears unlikely to modify the qualitative picture presented here. 

\section{Discussion and conclusions}\label{concs}
In this paper we have laid the theoretical groundwork for a comprehensive analysis of the turbulent (thermal) variability of tidal disruption event X-ray light curves. This work has built upon the analytical theories of \cite{MummeryBalbus22} and the numerical simulations of \cite{Turner23} and \cite{MummeryTurner24}. The key missing element of these earlier theoretical analyses was an ability to incorporate short-timescale correlations into the variability of the accretion models. These correlations  must be present in a physical system as the orbital timescale of the inner regions of a typical TDE disk is $\sim {\cal O}({\rm hours})$, comparable to the timescale over which observations are taken. 

The thermal variability of TDE disks is an area of study which has received little attention on the population level (while studies of individual objects of course exist), despite $f \gtrsim 40\%$ of optically discovered TDEs showing some X-ray emission \citep{Guolo24}, with the vast majority of these X-ray bright sources showing pure thermal spectra \citep[i.e., only 3 of the 17 X-ray bright TDEs in][have some coronal emission]{Guolo24}\footnote{TDEs discovered by X-ray surveys do appear to be much more likely to show coronal emission, but this seems likely to be a selection effect -- when TDEs do produce coronal emission (i.e., rarely), they typically become much more X-ray bright and therefore are more likely to be discovered by blind all-sky searches with an X-ray instrument.}.

In addition to developing this theory, we have performed the first analyses of the long-timescale variability of two TDEs, namely ASASSN-14li and AT2022lri. These systems were chosen both for their data quality, but also their spectral properties (thermal dominated) and differing variability structures. ASASSN-14li shows particularly simple stochastic variability, and is extremely well described by the theory developed here. ASASSN-14li has a well defined flux-RMS relationship, which spans over 3 orders of magnitude (in both the mean and standard deviation) of the luminosity. This is preliminary, though compelling, evidence therefore that tidal disruption events will join both X-ray binaries and AGN in showing this behavior. 

Where ASASSN-14li breaks with more typical accretion disk variability studies is in its actual luminosity distribution, which is clearly not log-normal (Figure \ref{fig:14li_data}, \ref{fig:14li_results}). Physically this can be traced to fact that these systems are observed in the Wien-tail of the disk spectrum, which exponentially suppresses the luminosity in {\it dimming} events, while only amplifying the luminosity with a power-law dependence for  {\it brightening} events. Statistically this introduces a pronounced asymmetry in the luminosity distribution, one that is clearly observed.  The properties of these light curve outliers can be understood within the framework of extreme value statistics (see Figure \ref{fig:evs} in section \ref{evs}). It is interesting to note therefore that the variability in the ultra-violet emission detected at late times in ASASSN-14li \citep[when the detected emission is coming from a disk, e.g.,][]{vanVelzen19, MumBalb20a, Mummery_et_al_2024} should be log-normally distributed, as this emission is well into the bulk of the disk spectrum and should therefore be symmetric. 

\begin{figure}
    \centering
    \includegraphics[width=0.49\linewidth]{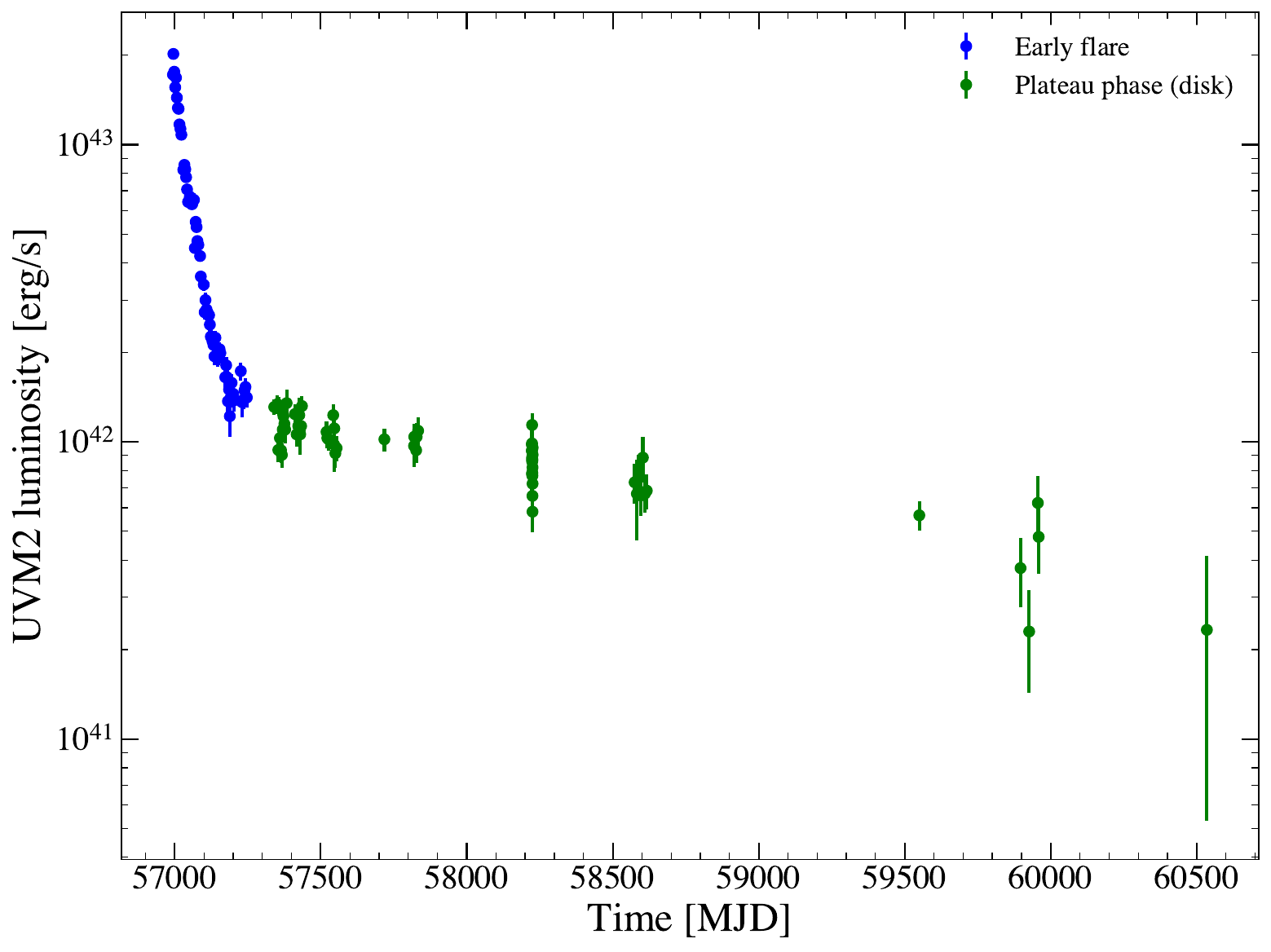}
    \includegraphics[width=0.49\linewidth]{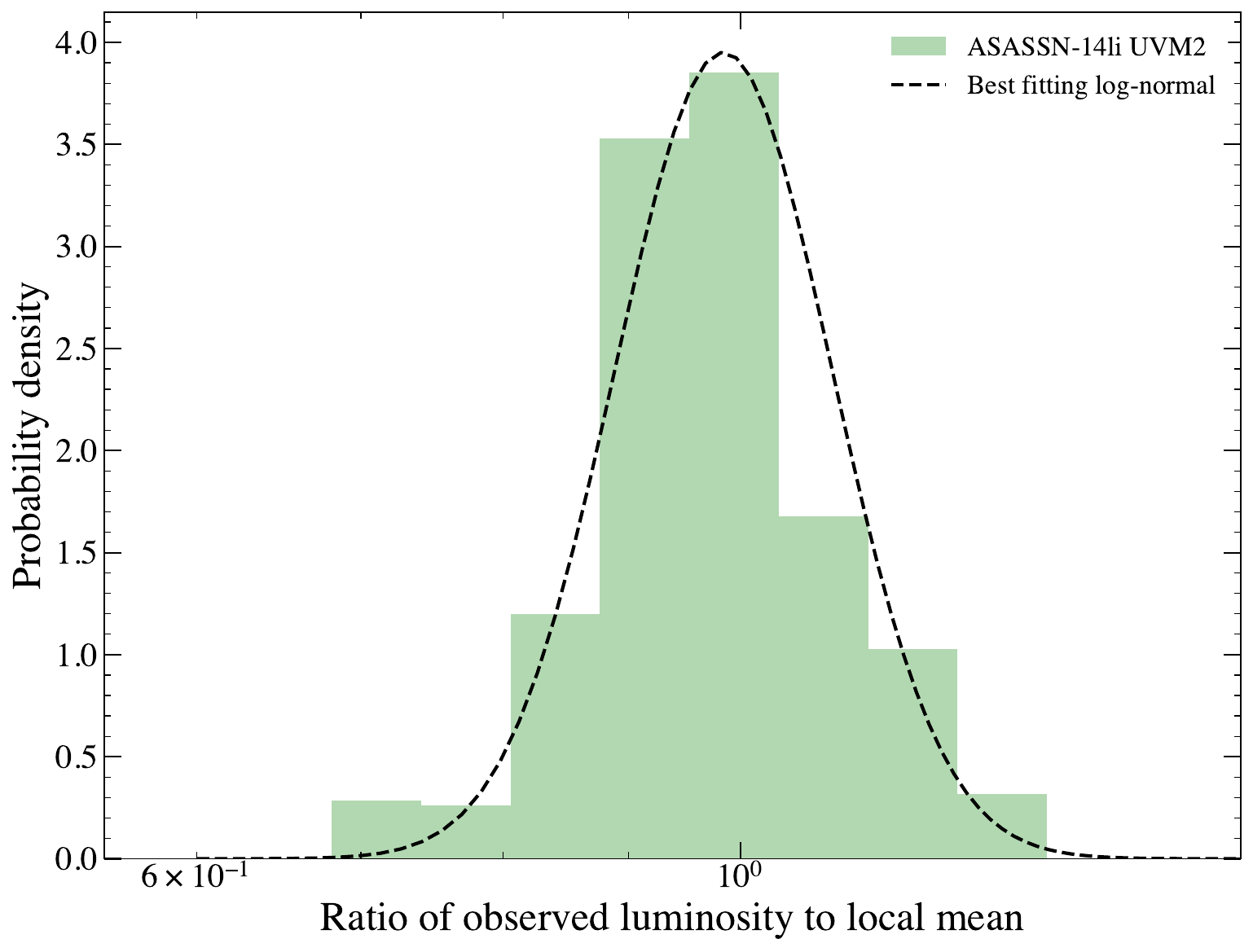}
    \caption{The variability in the UVM2 light curve of ASASSN-14li. On the left we show the {\it Swift} observations of ASASSN-14li over 3500 days, with the early flare denoted by blue points (non-disk emission), and the late time plateau (disk emission) denoted by green points. The fractional variability increases in the disk-dominated phase. This plateau-phase variability is well described by a log-normal distribution (right panel), a result of the flux originating from a wide range of radii in the disk and the observations not being in the Wien tail, which is what produces the asymmetry in the X-ray luminosity distribution.  }
    \label{fig:uv}
\end{figure}

This UV-luminosity distribution symmetry can be simply verified for ASASSN-14li, as it is one of the best studied TDEs at late times in the UV. In Figure \ref{fig:uv} we show the {\it Swift} UVM2 light curve of ASASSN-14li \citep[data taken from][]{GuoloMum24}, with the early flare denoted by blue points (non-disk emission), and the late time plateau (disk emission) denoted by green points. This plateau-phase variability is well described by a log-normal distribution (right panel), a result of the flux originating from a wide range of radii within the disk and the UV observations not being in the Wien tail, which is what produces the asymmetry in the X-ray luminosity distribution.

The source AT2022lri shows a more complex variability structure, with a pronounced change in fractional variability occurring roughly $\sim 60$ days beyond the first X-ray detection. Beyond this transitional time the variability is well described by the theory developed here, much like ASASSN-14li, and also shows a flux-RMS relationship which is near linear (despite not being described by a log-normal distribution, Figure \ref{fig:22lri_results}). Before this transition the variability is much higher, but still well described by the theory developed here. 

Many accreting systems are known to undergo transitions in variability structure, although this is usually associated with a corresponding change in spectral state \citep{Uttley11}. AT2022lri did not change spectral state, suggesting something unique to TDE disks may be happening here. We can foresee two plausible reasons why a TDE disk may undergo an extreme variability transition like this, without undergoing a spectral transition, namely 
\begin{enumerate}
    \item The disk was being initially fed in a highly time varying/clumpy manner. As a TDE disk is fed by returning debris from a tidally disrupted star, with typical returning timescale $t_{\rm fb} \sim {\cal O}({\rm months})$, it is possible that on initial timescales $t \lesssim t_{\rm fb}$ the feeding of the disk at the $\sim$ tidal radius will be highly time variable, which could filter down into a significantly more variable inner disk structure. The transition at $t \sim 60$ days could then correspond to a transition from variability dominated by this ``clumpy'' feedback to variability dominated by usual disk turbulence. 
    \item The disk at times $t \lesssim 60$ days was significantly thicker, and the disk then underwent a rapid thinning to a lower aspect ratio, and thereafter stayed with that new scale height. It seems likely that TDE disks will be thickest at early times (when the disk feeding is at its highest fraction of Eddington), before thinning down when the accretion rate drops at times beyond peak light. It has been demonstrated in numerical simulations \citep{Turner23} that the fractional variability of the disk temperature is a (near linear) function of disk scale height, owing to the larger turbulent eddies that can fit into a disk with a larger scale height. In a disk supported by radiation pressure, the disk aspect ratio is linearly proportional to the bolometric luminosity of the disk.  In this framework, all of the variability observed in AT2022lri would be simple turbulent variability of the temperature, and it is the change in scale height (i.e., luminosity) leading to a modified temperature variability which leads to the change in behavior at $\sim 60$ days. 
\end{enumerate}
Working out which (if either) of these scenarios is correct will likely require a more detailed analysis of the short timescale variability of AT2022lri than performed here. It seems likely that the correlation timescale in these two situations may be quite different, with the clumpy fallback model likely having some correlation in the temperature on the order of timescales set at the feeding radius (i.e., $\sim$ the tidal radius), which will necessarily be longer than those set at the innermost radii (which would set the timescale in the second, thicker disk, scenario). Indeed, it can be rigorously proven that variability on timescales shorter than the viscous time at the feeding radius is exponentially suppressed  in this clumpy fallback scenario \citep[this is merely a property of the diffusive nature of the accretion process][]{Mummery23b}. The fact that $\sim$ order of magnitude dimming events where observed on $\sim$ hour timescales for this source \citep{Yuhan24} strongly favors the second, thinning-disk, scenario. 

The study of short timescale (likely intra-observation) variability of TDEs will offer a fascinating, and possibly unique, probe into the fundamental physics of short-timescale accretion disk turbulence and variability.  This is precisely because their natural evolutionary timescales span particularly interesting observational windows, and the fact that TDEs appear to have particularly ``clean'' spectra. The orbital timescales in their innermost regions are of order $\sim {\cal O}({\rm hours})$, which we have argued sets the absolute lower limit over which variability may be expected to be observed, and is exactly the timescale over which a typical X-ray observation is taken. The global evolution of these systems takes place over $\sim {\cal O}({\rm weeks-months})$, again a very observationally convenient timescale to probe fundamental questions of disk physics. The fact that the spectra of these systems also appears to be dominated by a relatively well understood (in an average sense) component, will allow for the isolating of variability features with minimal complications.  Important questions that may be answered with future observational campaigns dedicated to studying the short timescale turbulent variability of TDE disks include 
\begin{itemize}
    \item What is the {\it shortest} timescale over which there is observable variability in the  temperatures of TDE accretion disk systems? And how is this related to the orbital timescale of the inner disk regions? This is ultimately a fundamental question regarding the timescales governing turbulent fluctuations in accretion flows, and will have broad implications. 
    \item Does the short timescale variability in TDE X-ray light curves show the predicted energy dependence derived here for all TDE sources?  In other words, is the variability in the X-ray light curves well described by an intrinsic process (a stochastic disk temperature profile), or is it set by some other external process? 
\end{itemize}
On longer timescales a range of other important questions can be posed
\begin{itemize}
    \item Are pronounced dimming events more common than brightening events across the entire TDE population? Or, do all sources show the asymmetric luminosity distributions seen in ASASSN-14li and AT2022lri? 
    \item Do all TDEs show a long-timescale flux-RMS relationship similar to those that we have shown to exist in ASASSN-14li and AT2022lri? 
    \item How is variability structure related to spectral state, and do TDE disk systems also show a more variable state when they transition to a coronal dominated spectrum (a common feature of X-ray binary disks)? 
\end{itemize}
It is the hope of the author that the stochastic light curve model developed here will facilitate this study, and provide a means to quantify the ``typical'' variability expected from disk turbulence. 

Another use of this model will be in quantifying when an observed variability is ``untypical'', in some meaningful sense. Within the TDE literature there have been claims of quasi-periodicity in the variability of early time X-ray light curves of TDEs \citep[e.g.,][]{Pasham24}. To make a statistical claim of quasi-periodicity one must of course perform a null hypothesis test, and compute how common it would be to see, for example, $N$ flares separated by roughly the same time interval in a random light curve. Conventionally this null hypothesis test is performed by sampling light curves from a given power spectrum (normally red or white noise), and determining how many of those light curves show similar statistical periodicity features. We have stressed in this work that 
it is inevitable that accretion flows will have temporal correlations on a range of different timescales (with different correlation scales ranging from the orbital timescale of the innermost regions up to the evolutionary timescale at the feeding i.e., $\sim$ tidal, radius of the flow). It certainly appears plausible that having a correlation timescale of order $t_{\rm corr} \sim$ a few days could lead to an increased probability of seeing flares separated by $\Delta t \sim t_{\rm corr} \sim$ a few days in a restricted observing window. One could use the correlated light curve generator which we derive here as an alternate null hypothesis test, by sampling large numbers of stochastic light curves with different correlation times, fractional temperature variabilities (etc.), and computing how many of those light curves show quasi-periodic variability structures within a limited time window. 

We stress that the point of this discussion is not that any claimed quasi-periodicity is incorrect (it is certainly physically plausible that a TDE could precess in the early phases of its evolution due to the Lens-Thirring effect, and this could induce an associated quasi-periodicity in the X-ray emission), just that future analyses could make use of a modified, physically-motivated, null hypothesis test when making such claims. 

To conclude, in this paper we have developed a stochastic light curve model for evolving TDE disks. This model incorporates short-timescale correlations which must be present in a turbulent accretion flow, the effects of observing TDE systems in the Wien-tail which exponentially enhances variability, the long-timescale evolutionary impacts of the disk having a finite initial mass budget, and a physically motivated log-normal distribution for the disk temperature. 

We have demonstrated, using extreme value statistics, that extreme {\it dimming} events are more likely to be observed from a TDE system than {\it brightening} events, and therefore that TDE X-ray luminosity distributions should be asymmetric on a logarithmic scale \citep[see also][]{MummeryBalbus22}. We have shown that the short timescale variability observed in TDEs will offer a powerful probe of the physical timescales over which accretion flows vary, which will offer profound insights into fundamental questions of disk physics. This simply results from a coincidence that the orbital time at the innermost stable circular orbit of a $\sim 10^6 M_\odot$ black hole is comparable to an order $\sim {\cal O}(0.1)$ fraction of the typical observing time of a given XMM observation. The short (intra-observation) timescale variability seen in ASASSN-14li and AT2022lri is strongly supportive of the theory developed here. 

On longer, global evolutionary, timescales the model developed here provides a good description of the variability observed from two well-observed TDE systems, ASASSN-14li and AT2022lri. These systems show characteristic hallmarks of the theory developed here,  namely large amplitude dimming events (AT2022lri), asymmetric luminosity distributions (ASASSN-14li and AT2022lri), and a near-linear RMS-flux relationship (ASASSN-14li and AT2022lri). Future, comprehensive, analyses of the full TDE population will confirm whether or not these features are universal to X-ray bright TDE disks.

\section*{Acknowledgments} 
I am grateful to Francesco Mori for introducing me to the fascinating subject of extreme value statistics, and for various helpful conversations.  I am grateful to Yuhan Yao and Muryel Guolo for sharing their long-timescale X-ray data, and for helpful discussion. I am particularly grateful to Guolo for providing additional short-timescale data for both sources, without which section 5.3 and Figure 10 would not have been possible. Discussions with Megan Masterston, Erin Kara and Joheen Chakraborty were extremely illuminating, and motivated much of this work.  This work was supported by a Leverhulme Trust International Professorship grant [number LIP-202-014]. For the purpose of Open Access, AM has applied a CC BY public copyright license to any Author Accepted Manuscript version arising from this submission. The theory developed in this paper benefited from insight from numerical simulations run in \citep{MummeryStone24}, which used resources of the National Energy Research Scientific Computing Center (NERSC), a Department of Energy Office of Science User Facility using NERSC award FES-ERCAPm4307, and \citep{MummeryTurner24} which {used resources provided by the Cambridge Service for Data Driven Discovery (CSD3) operated by the University of Cambridge Research Computing Service (\url{www.csd3.cam.ac.uk}), provided by Dell EMC and Intel using Tier-2 funding from the Engineering and Physical Sciences Research Council (capital grant EP/T022159/1), and DiRAC funding from the Science and Technology Facilities Council (\url{www.dirac.ac.uk}). }
 
\section*{Data accessibility statement}
All observational data used in support of this manuscript is publicly available through the High Energy Astrophysics Science Archive Research Center (HEASARC, \href{https://heasarc.gsfc.nasa.gov}{https://heasarc.gsfc.nasa.gov}). The stochastic light curve model developed in this paper can be downloaded from \href{here}{link will go here}.

\bibliographystyle{mnras}
\bibliography{andy}

\appendix
\section{The X-ray luminosity cumulative distribution function and its inverse}\label{CDF}
We explicitly require an expression for the cumulative distribution function $\Phi_L$ of the X-ray luminosity distribution $p_L$ and its inverse, denoted $Q_L$ (for ``quartile''). This can be done analytically following the ``law of the unconscious statistician'', which states that for two variables related by a monotonic function $f$ (i.e., $y = f(x)$ with $f$ either always increasing or always decreasing) we have $p_Y(y)\,{\rm d}y = p_X(x)\,{\rm d}x$. We know that the (normalised) X-ray luminosity  $y \equiv L/L_0$ is related to the (normalised) disk temperature $x \equiv k_B f_{\rm col}T_p / E_l$ by such a function $y = x^\eta \exp(-1/x) \equiv g(x)$, and that $x$ is log-normally distributed, i.e., $z \equiv \ln(x)$ is normally distributed with some mean $\mu_N$ and variance $\sigma_N^2$. The function $g(x)$ is a monotonically increasing function of its argument, and therefore 
\begin{equation}
    \Phi_L(\ell) = \int_0^\ell p_L(\ell')\, {\rm d}\ell' = \int_0^{\ell/L_0} p_Y(y')\, {\rm d}y' = \int_0^{g^{-1}(\ell/L_0)} p_X(x')\, {\rm d}x' ,
\end{equation}
where $g^{-1}(y)$ is the inverse of $g(x)$. As the natural logarithm is also a monotonically increasing function of its argument, we can also state that 
\begin{equation}
    \Phi_L(\ell) = \int_0^{g^{-1}(\ell/L_0)} p_X(x')\, {\rm d}x' = \int_{-\infty}^{\ln(g^{-1}(\ell/L_0))} p_Z(z')\, {\rm d}z' .
\end{equation}
Note however that this final integral is just the cumulative distribution function of the normal distribution, and so equals
\begin{equation}\label{PHIL}
    \Phi_L(\ell) =  \int_{-\infty}^{\ln(g^{-1}(\ell/L_0))} p_Z(z')\, {\rm d}z' = \Phi_N(\ln(g^{-1}(\ell/L_0))) = {1\over 2}\left[1 + {\rm erf}\left( {\ln(g^{-1}(\ell/L_0)) - \mu_N\over \sqrt{2}\sigma_N}\right) \right] ,
\end{equation}
where ${\rm erf}(z)$ is the usual error function
\begin{equation}
    {\rm erf}(z) \equiv {2\over \sqrt{\pi}} \int_0^z e^{-t^2} \, {\rm d}t. 
\end{equation}
As $y = g(x)$ has a well defined inverse \citep{MummeryBalbus22}
\begin{equation}
    g^{-1}(y) = 1/\left(\eta W\left[y^{1/\eta}/\eta\right]\right) ,
\end{equation}
where $W$ is the Lambert $W$ function \citep{Corless96}, this is a fully analytical solution for the cumulative distribution function of the X-ray luminosity. This result is the one used in the main body of the paper. We remind the reader that the mean and variance of the underlying normal distribution are related to the (more observable) parameters of the temperature distribution by 
\begin{align}
    \mu_N &= \ln\left({k_B\mu_T \over E_l }{\mu_T \over \sqrt{\mu_T^2 + \sigma_T^2}} \right), \\ 
    \sigma_N^2 &= \ln\left(1 + {\sigma_T^2 \over \mu_T^2 }\right). 
\end{align}
We also require the inverse cumulative distribution function (sometimes referred to as the $Q$-function), which we denote 
\begin{equation}
    \ell = Q_L(\Phi_L) ,
\end{equation}
where $0 \leq \Phi_L \leq 1$. Inverting equation \ref{PHIL}, we find 
\begin{equation}\label{QL}
    \ell = Q_L(\Phi_L) = L_0 \, g\left(\exp\left[\mu_N + \sqrt{2} \sigma_N\,  {\rm erf}^{-1}\left(2\Phi_L - 1\right)\right]\right),
\end{equation}
where ${\rm erf}^{-1}(z)$ is the inverse error function.  

\section{Taking the limit $\sigma_T/\mu_T \to 0$ }\label{limproof}
Various proofs that in the limit $\sigma_T/\mu_T \to 0$ we have $p_L(l) \to \delta(l-\overline L)$ can be constructed. In this Appendix we provide the simplest. Starting with the above definitions of $\mu_N$and $\sigma_N$ we see that 
\begin{align}
    \lim_{\sigma_T/\mu_T \to 0} \mu_N &\to \ln\left({k \mu_T \over E_l} \right), \\
    \lim_{\sigma_T/\mu_T \to 0} \sigma_N &\to \left({\sigma_T \over \mu_T}\right)^2 \to 0. 
\end{align}
Then from equation \ref{QL} we see that when we sample from a distribution with these properties we always find 
\begin{equation}
    \lim_{\sigma_T/\mu_T \to 0} \left[l \right]=\lim_{\sigma_T/\mu_T \to 0} \left[Q_L(\Phi_L) \right] \to L_0 \, g\left(\exp\left[\mu_N\right] \right) , 
\end{equation}
independent of the value of the random number $\Phi_L$ one samples. In this limit this equals 
\begin{equation}
    l = L_0 \, g\left(\exp\left[\mu_N\right] \right) = L_0 \left({k \mu_T \over E_l} \right)^\eta \exp\left(-{E_l \over{k \mu_T } }\right) = \overline L , 
\end{equation}
which is exactly the thin disk theory prediction for the mean evolution. The fact that irrespective of the random number $\Phi_L$ one samples one always returns a value $\overline L$ from a probability density function implies that that probability density function is equal to a delta function, i.e., 
\begin{equation}
    \lim_{\sigma_T/\mu_T \to 0} p_L(l) \to \delta(l - \overline L). 
\end{equation}

\section{The importance of the Wien tail}\label{app:C}
\begin{figure}
    \centering
    \includegraphics[width=0.65\linewidth]{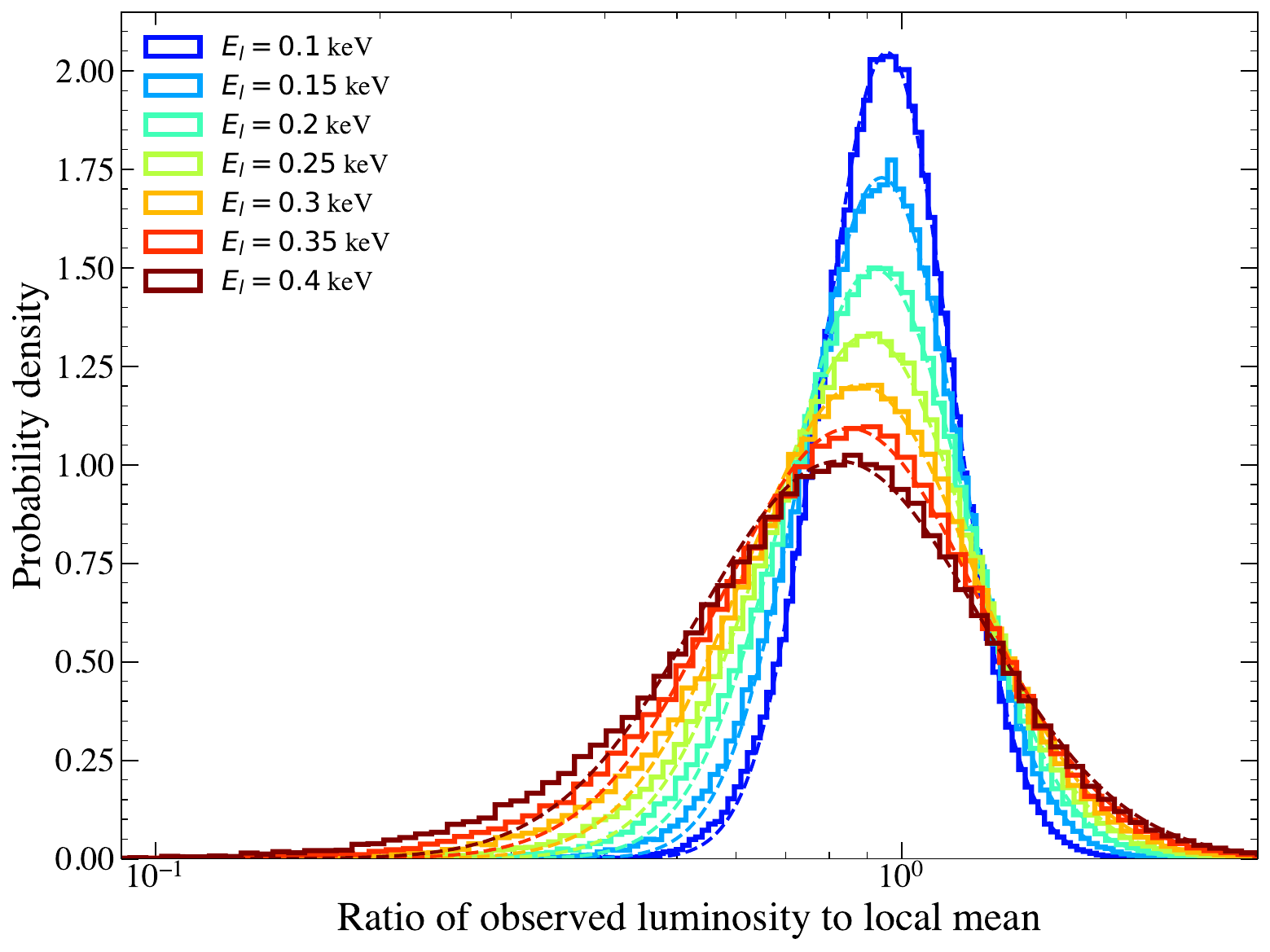}
    \includegraphics[width=0.65\linewidth]{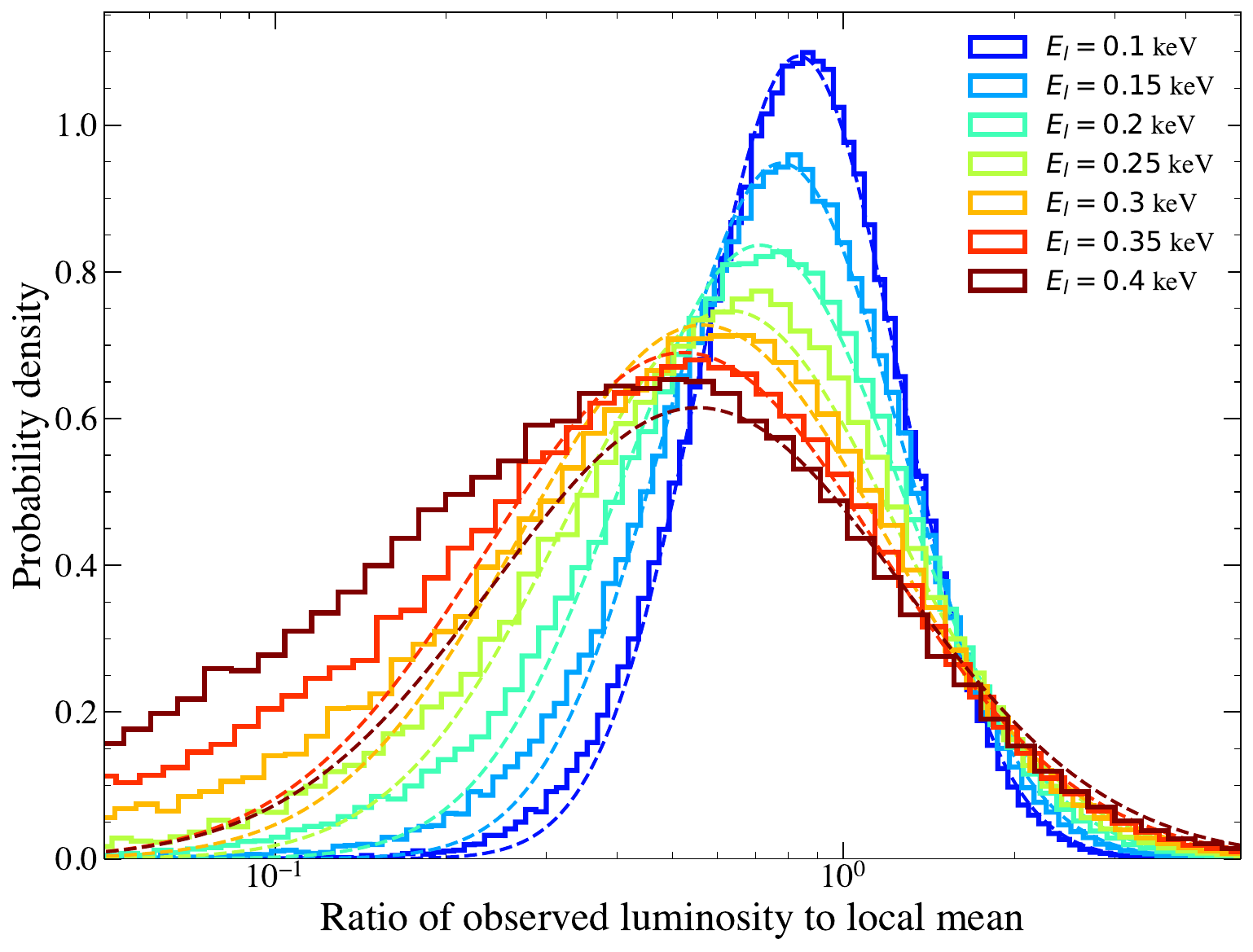}
    \caption{The importance of observing TDEs in the Wien tail. Both tidal disruption events analysed in this work showed a pronounced asymmetry in their luminosity distributions, which we argued was a result of the observations being taken in the Wien tail of the disk spectrum. This can be understood on theoretical grounds, but in this Figure we demonstrate it explicitly with Monte Carlo simulations. In this Figure we produce Monte Carlo distributions of ${\cal R}$ -- the ratio of the observed luminosity to its local mean.  The observing cadence and mean temperature evolution are set equal to that of ASASSN-14li, and all that is varied is the fractional temperature variability (upper vs lower panels) and the observing energy (coloured curves). In the upper plot we take $\sigma_T/\mu_T = 0.06$ (i.e., precisely the value used in the main body of the text), while in the lower panel we take $\sigma_T/\mu_T = 0.12$. By reducing the observing energy (and therefore putting more of the disk luminosity in the observing band, and less in that observers Wien tail), we see that the distributions of ${\cal R}(E)$ become increasingly well described by a log-normal distribution (dashed curves), irrespective of the fractional temperature variability. The fact that we observe an asymmetry in real TDE luminosity distributions simply highlights the fact we are observing these systems in the Wien-tail and only seeing a small fraction of the bolometric luminosity of the disk.   }
    \label{fig:wien}
\end{figure}

In this Appendix we demonstrate numerically a fact we have stressed throughout this paper, that it is the process of observing TDEs in their Wien tail which produces a luminosity distribution asymmetry. In Figure \ref{fig:wien} we produce Monte Carlo distributions of ${\cal R}$ -- the ratio of the observed luminosity to its local mean, in an identical fashion to what was done for ASASSN-14li and AT2022lri in the main body of the paper.  In this Figure the observing cadence and mean temperature evolution are set equal to that of ASASSN-14li, and all that is varied is the fractional temperature variability (upper vs lower panels) and the observing energy (coloured curves). In the upper plot we take $\sigma_T/\mu_T = 0.06$ (i.e., precisely the value used in the main body of the text), while in the lower panel we take $\sigma_T/\mu_T = 0.12$. By reducing the observing energy (and therefore artificially putting more of the disk luminosity in the observing band, and less in the observers Wien tail), we see that the distributions of ${\cal R}(E)$ become increasingly well described by a log-normal distribution (dashed curves), irrespective of the fractional temperature variability. The fact that we observe an asymmetry in real TDE luminosity distributions simply highlights the fact we are observing these systems in the Wien-tail and only seeing a small fraction of the bolometric luminosity of the disk.

\label{lastpage}

\end{document}